\newcommand{\rev}[1]{#1}

\newcommand{\Mjup}{$\mathrm{M_{Jup}}$}
\documentclass[preprint2]{pp7}
\usepackage{xcolor}
\usepackage{gensymb}
\usepackage[flushleft]{threeparttable}
\usepackage[switch]{lineno}

\usepackage[normalem]{ulem}

\input{pp7.h}

\begin{document}

\title{\textbf{\LARGE Optical and Near-infrared View\\ of Planet-forming Disks and Protoplanets}}

\author {\textbf{\large Myriam Benisty}}
\affil{\small\it Univ. Grenoble Alpes, CNRS, IPAG, 38000 Grenoble, France}
\author {\textbf{\large Carsten Dominik}}
\affil{\small\it Anton Pannekoek Institute for Astronomy, University of Amsterdam, Science Park 904, 1098 XH Amsterdam, The Netherlands}
\author {\textbf{\large Katherine Follette}}
\affil{\small\it Physics \& Astronomy Department, Amherst College, 21 Merrill Science Dr., Amherst, MA 01002}
\author {\textbf{\large Antonio Garufi}}
\affil{\small\it INAF, Osservatorio Astrofisico di Arcetri, Largo Enrico Fermi 5, I-50125 Firenze, Italy}
\author {\textbf{\large Christian Ginski}}
\affil{\small\it Anton Pannekoek Institute for Astronomy, University of Amsterdam, Science Park 904, 1098 XH Amsterdam, The Netherlands}
\author {\textbf{\large Jun Hashimoto}}
\affil{\small\it Astrobiology Center, National Institutes of Natural Sciences, 2-21-1 Osawa, Mitaka, Tokyo 181-8588, Japan}
\author {\textbf{\large Miriam Keppler}}
\affil{\small\it Max Planck Institute for Astronomy, Königstuhl 17, 69117, Heidelberg, Germany}
\author {\textbf{\large Willy Kley}}
\affil{\small\it Institut für Astronomie und Astrophysik, Universität Tübingen, Auf
der Morgenstelle 10, 72076 Tübingen, Germany}
\author {\textbf{\large John Monnier}}
\affil{\small\it Astronomy Department, University of Michigan, Ann Arbor, MI 48109, USA}

\begin{abstract}
\baselineskip = 11pt
\leftskip = 1.5cm 
\rightskip = 1.5cm
\parindent=1pc
{\small 
In this chapter, we review the breakthrough progress that has been made in the field of \textit{high-resolution}, \textit{high-contrast} \textit{\rev{optical} and near-infrared} imaging of planet-forming disks. These advancements include the \textit{direct detection} of protoplanets embedded in some disks, and \textit{derived limits} on planetary masses in others. Morphological substructures, including: rings, spirals, arcs, and shadows, are seen in all imaged infrared-bright disks to date, and are ubiquitous across spectral types. These substructures are believed to be the result of disk evolution processes, and in particular disk-planet interactions. Since small dust grains that scatter light are tightly bound to the disk's gas, these observations closely trace disk structures predicted by hydrodynamical models and serve as observational tests of the predictions of planet formation theories. We argue that the results of current and next-generation high-contrast imaging surveys will, when combined with complementary data from ALMA, lead to a much deeper understanding of the co-evolution of disks and planets, and the mechanisms by which planets form.
 \\~\\~\\~}
\end{abstract}  

\section*{In memory of Prof. Willy Kley}
\noindent \textit{We dedicate this chapter to our co-author Willy Kley, who unexpectedly passed away in December 2021. His unique contributions to our field, and his optimism and kindness will be remembered for a long time. An obituary written by Richard Nelson can be found at the end of this Chapter.}

\section{Introduction} \label{sec:intro}
\subsection{Motivation}
Transit, radial velocity, and imaging surveys have discovered and confirmed over 4500 exoplanets\footnote{September 2021; \url{https://exoplanets.nasa.gov/discovery/exoplanet-catalog/}} indicating that planet formation is a robust and efficient process. Observing campaigns have also confirmed that planetary systems show an incredible diversity in the nature of individual planets and in the architecture of planetary systems. Confirmed planetary systems span a range of star-planet separations and masses, from widely-separated Super-Jupiters at tens to hundreds of astronomical units (au) \citep[e.g.][]{Marois2008} to compact Super-Earth systems compressed to within just a few au from their host stars \citep[e.g.][]{Lissauer2011}. Planet formation has also been confirmed to occur in multiple star systems, where around a dozen planets have been found \citep{Kostov2016}. The origin of this extreme diversity, and whether it is inherited from the earliest stages of planet formation in protoplanetary disks, is unknown.  Since protoplanetary disks set the initial conditions for planet formation, the observed diversity of exoplanets might well be related to the diversity of disk physical properties. The evolution of bulk disk gas and dust will be key in determining where, when, and how planets form and evolve. The interaction of planets with their host disk is another key piece of the puzzle in planetary evolution; massive planets form first and can dramatically affect disk structure \citep{Bae2018}, influencing the formation and evolution of lower-mass planets, and dominating the dynamical evolution of the planet-star-disk system (see Chapters \textit{Bae et al.}; \textit{Paardekooper et al.}). 

To understand the diverse outcomes of planet formation, it is therefore crucial to understand the structure of protoplanetary disks from the epoch of formation and throughout the disk evolutionary sequence, using the disks' morphological appearances are essential clues. Planet formation and disk evolution processes occur simultaneously, influence one another, and jointly shape disk structures. The two main imprints of these processes on the disk are ($i$) perturbations in the disk structure (that is, formation of substructures), and ($ii$) spatial differentiation of gas and dust.  High-resolution imaging of gas and dust grains in protoplanetary disks affords a direct probe of both imprints. To probe disk evolution, large samples of high-resolution multiwavelength disk imagery accross a wide range of disk evolutionary stages are required.    

Until recently, disk observations were limited in angular resolution and could not provide clear constraints on the physical processes driving their evolution. Modeling spectral energy distributions (SEDs) derived from whole-disk photometry cannot unambiguously determine the disk structure, as it is highly degenerate \citep{Woitke2019}. \rev{In the past few years,} major progress in instrumentation has enabled some limitations to be overcome and has dramatically affected both observational and theoretical studies of planet formation. The  advent of the Atacama Large Millimeter/sub-millimeter Array (ALMA) allowed us to spatially resolve the thermal emission of cold, large (mm-sized) dust grains and molecules in circumstellar disks \citep[e.g.,][]{Dutrey2014} and circumplanetary disks (CPDs) \citep{Isella2019,Benisty2021}. In the optical and near-infrared \rev{(near-IR)} regime, telescopes with extreme adaptive optics (AO) enabled nearly diffraction-limited imaging ($\sim$4 au at 100\,pc), providing access to both thermal emission from young planets and dust in CPDs, as well as to the scattered light from small (micron-sized) dust grains \rev{in the disk surface layers} \citep{Keppler2018}. The first images of disks obtained at high resolution revealed a greater complexity than previously envisioned \citep{Quanz2011, hltau}, revolutionizing our view of protoplanetary disk evolution. The first surveys with ALMA and high-contrast near-IR imagers suggested that small-scale substructures are \rev{frequent} and have a variety of morphologies, locations, and properties \citep{Andrews2020, Garufi2018}. The direct comparison of images obtained at different wavelengths allows us to constrain the radial and vertical distributions of gas and dust in disks. Indeed, large grains settle to the midplane, and small grains are well coupled to the gas, covering the full vertical gas extent of the disk. Therefore, multi-wavelength observations trace different disk layers \rev{\citep[e.g.,][]{Dong2012b}} and enable construction of a three-dimensional view of the disk physical conditions.

One of the major \rev{quests} in our field is to identify and constrain favorable conditions for planet formation. \rev{We need to know where, when, and how planets form, and whether there are significant differences in the planet population as a function of system age and location in the disk.} \rev{In the core-accretion model \citep{Pollack1996}, planet formation starts with the growth of dust grains to planetesimal sizes.}  However, pressure gradients in the gas will lead to sub- or sometimes super-Keplerian velocities of the gas.  Dust grains then feel a drag force that causes them to drift radially towards regions of the disk with higher pressure, usually toward the inner disk. In \rev{the} absence of a way to reduce or stop radial drift, this effect results in all dust grain orbits decaying toward the \rev{star} before any planet could form \citep{Weidenschilling1977}. Local variations of the gas pressure (reversing the radial pressure gradient) provide a way to maintain these dust grains in the disk, as dust grains with Stokes number close to 1 \citep{Birnstiel2010} will move to the nearby local pressure maximum and get trapped there \citep[e.g.,][]{Pinilla2012}. In these so-called dust traps, grains can grow efficiently to pebble sizes, and when a high dust to gas ratio is reached, streaming instabilities can occur and lead to the formation of planetesimals \citep{youdin05}. These local variations of the gas pressure will lead to distinct observable features (substructures) at different tracers \rev{(gas lines, scattered light and thermal emission images)} \citep[e.g.,][]{Baruteau2019}. The typical size gas substructures is the pressure scale height, which is about one to several au at separations of few tens of au from the central star, a scale accessible with high angular resolution observations. 

Generally speaking, observed disk substructures can thus be both the \textit{cause} (as favored locations of planetesimal growth), and the \textit{consequence} (as imprints of massive forming planets) of planets in disks. Therefore, theoretical studies aimed at constraining the origin of substructures are pivotal (see Chapters \textit{Bae et al.}; \textit{Lesur et al.}; \textit{Paaderkooper et al.}).  In short, substructures can be created by the dynamical evolution of the gas in the disk, through (magneto-)hydrodynamical instabilities and disk winds \citep[e.g.,][]{Riols2020}. Such processes lead to both symmetric and asymmetric pressure perturbations, which will in turn trap dust grains. On the other hand, substructures may be signposts of embedded, yet-undetected, protoplanets. For example, planets launch spiral density waves at Lindblad resonances, deplete their orbit of material \citep{Kley2001} leading to dust and gas depleted gaps \citep{Bae2017}, or even large cavities. The characteristics of these gaps and cavities depend on the planet mass and orbital parameters and on the disk viscosity. Planet-disk interactions also often trigger large-scale hydrodynamical instabilities  \citep[vortices; ][]{Meheut2010} that lead to strong azimuthal asymmetries. Disk substructures can therefore serve as a guide to detect hidden planets \citep{Dong2019} as well as a local probe of the disk conditions while dynamical interactions are ongoing.  An extensive study of the prevalence and properties of substructures will allow us to determine the relative importance of the vaious mechanisms driving disk evolution, and can help constraining a yet-undetected population of exoplanets that can then be compared to the demographics of more evolved exoplanetary systems (see Chapter \textit{Bae et al.}). 

\rev{Early results from scattered light disk imaging were previously discussed in the Protostars \& Planets (PP) series (PPV; \citealt{Watson2007} and PPVI; \citealt{PPVIEspaillat}), and are summarized in the following paragraph. Hubble Space Telescope images provided the first evidence of substructures in disks, in particular around Herbig AeBe stars which presented extended nebulosity \citep{Grady1999, Fukagawa2004}. Spiral features were detected around HD100546 \citep{Pantin2000, Grady2001} and HD141569A \citep{Mouillet2001, Clampin2003}, and radial clearing and fading of the outer disk were observed in HD141569A and TW Hya respectively \citep{Weinberger1999,Krist2000}. Observations of the silhouette of edge-on disks enabled determination of disk scale heights \citep[e.g.,][]{Perrin2006}, supported by model fitting \citep{Watson2004}. Detailed analysis of multi-wavelength images of GG Tau provided evidence for dust stratification in the disk from the scattering phase function \citep{McCabe2002,Duchene2004}.  Between PPV and PPVI, near-IR polarimetric disk observations were conducted, in particular through the SEEDS (Strategic Explorations of Exoplanets and Disks with Subaru) survey \citep[e.g.,][]{Thalmann2010, Hashimoto2012, Mayama2012, Takami2013}, and  VLT/NACO observations \citep[e.g.,][]{Quanz2011, Quanz2012} providing further support for the presence of substructures (cavities, rings, spirals) in IR-bright disks. The focus of the \citet{PPVIEspaillat} PPVI review was on transition disks (there referred to as transitional, and pre-transitional disks).}

\begin{figure*}[!t]
\centering
  \includegraphics[width=0.9\textwidth]{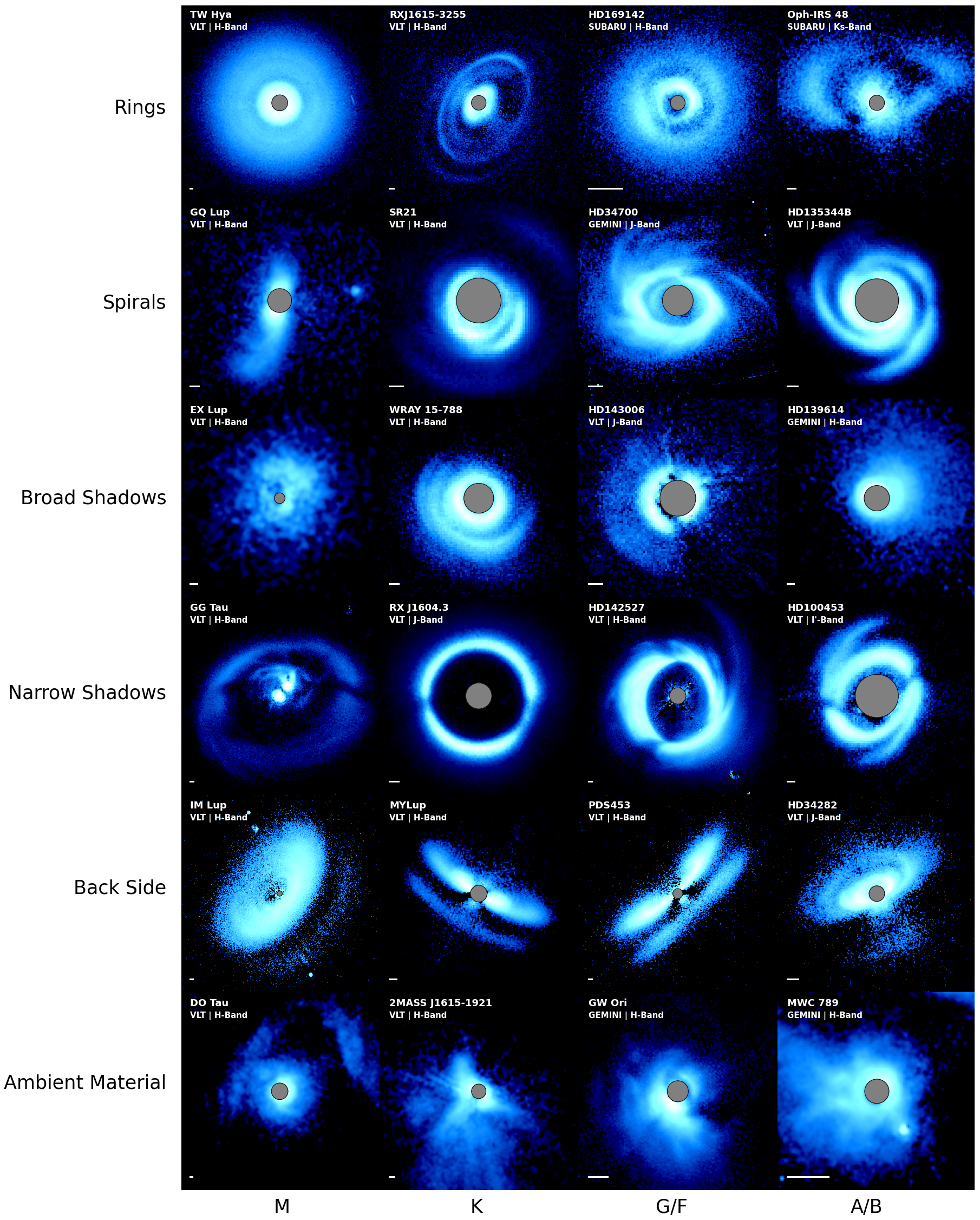}
  \caption{\small Gallery of disks observed with the SPHERE, GPI, and HiCIAO instruments, highlighting the wide range of morphological features present in planet-forming disks and their ubiquity across spectral types. All images are normalized within the depicted stamp and colorbars are arcsinh scaled. A 50\,au scalebar appears in the lower left corner of each image, and grey central circles mark regions obscured by a coronagraph. North is up and East to the left in all images. The star name, instrument, and wavelength are indicated in the upper left corner of each image. In all cases, the polarimetric Q$_\phi$ image is shown, with the exception of GG\,Tau, for which we show polarized intensity.  Images are coarsely sorted by spectral type (x axis) and morphological type (y axis), though we note that many disks exhibit features matching multiple morphological categories. \rev{All relevant references for individual systems are listed in appendix~\ref{App: ref: main gallery}. Figures \ref{fig:alma}, \ref{fig:spiral-gallery} and \ref{fig:shadows} provide additional disk images.} }
  \label{fig:gallery}
 \end{figure*}

\rev{Since PPVI}, monumental progress has been achieved in the field of high-resolution disk imaging, indicating the \rev{ubiquity} of substructures in scattered light. 
Another major result since PPVI, is the first and so far unique discovery of a protoplanetary system still embedded in its birth environment, PDS\,70.  
This chapter is a review of major results in the field of optical and near-IR high-resolution disk imaging, as well as highlights of relevant results from other, highly complementary, observational techniques. \rev{Large emphasis is placed on the disk substructures resolved by polarimetric images from ground-based telescopes since much of the recet work of the community has been in this area. Figure~\ref{fig:gallery} is a compilation of such images, highlighting a number of the results reviewed in this chapter. The data presented in the figure presents the pinnacle of our current observational capabilities in the optical and near-IR direct imaging of circumstellar disks, achieved thanks to leading observational facilities such as the Subaru telescope, the Gemini-South telescope, and the ESO Very Large Telescope. The methodology and instruments used are described} in Section \ref{sec:method} of the chapter. Section \ref{sec:theory} provides a general theoretical background for understanding substructures, Section \ref{sec:globalstructure} reviews results on the global structure of disks and the properties of dust grains, and Section \ref{sec:main_substructure} presents an overview of observations of disk substructures, such as cavities, rings, spiral arms, and shadows. In Section \ref{sec:main_protoplanet}, \rev{we discuss the discovery of the first young embedded planetary system along with a review of current upper limits for other systems}. We discuss future avenues for research in the field of direct imaging of disks and protoplanets in Section \ref{sec:future}, and \rev{summarize our chapter in Section \ref{sec:summary}}. 

\section{Methodology} \label{sec:method}

At optical and near-IR wavelengths, the star itself and the hot innermost disk regions are the primary sources of photons. While the current generation of single-dish telescopes cannot spatially resolve thermal emission from the disk regions inwards of a few au, thermal emission from planets located at separations larger than 10 au are, in principle, observable \rev{if these planets are not heavily embedded}. At similar separations, 10s of au from the star, the protoplanetary disk can be mapped in scattered light. Small (sub-)$\mu$m-sized dust grains and aggregates, which are well coupled to the gas, can efficiently scatter stellar photons at optical and near-IR wavelengths. \rev{The gas- and dust-rich planet-forming disks depicted in Figure\,\ref{fig:gallery} are optically thick to optical and near-IR radiation \citep{Chiang1997}}. Scattered light observations will therefore be sensitive to the optical depth $\tau\sim$ 1 layer located in the surface of the disk. As such, it will be directly related to \rev{both} the stellar irradiation (that drops off radially with the squared distance to the star) and the local inhomogeneities in the surface. \rev{Therefore, the net amount of observed scattered photons depends on both the number of photons incident on the disk surface and the availability of scattering dust grains. } 

\subsection{Requirements}
\rev{High-contrast} imaging of planet-forming disks \rev{in the optical and near-IR} faces several challenges, namely the need for high angular resolution and high contrast, as well as cancellation of atmospheric perturbations if observations are obtained with ground-based instruments. 
The structures that we aim to observe have typical widths of the pressure scale height $H$ in the gas \citep{Dullemond2018}, which depends on the local sound speed and on the Keplerian angular velocity. For typical disk models, $H/R$ is typically 0.05-0.1 at radii of tens of au, implying that substructures have typical widths of a few au at most. This requires the use of large diameter telescope facilities. Indeed, nearby reservoirs of young stellar objects in low-mass star forming regions (e.g., the Taurus and Lupus molecular cloud complexes, or the Chamaeleon I cloud) are located at distances between $\sim$150\,pc and 200\,pc (\citealt{Dzib2018, Luhman2018}). At such distances, 6-8\,au translates into on-sky angular scales of $\sim$0.040\arcsec{}, for a diffraction limited resolution element on an 8m-class telescope in the H-band ($\lambda$ = 1.65 $\mu$m). This also means that, with the largest currently available telescope facilities (8-10\,m telescopes such as the Very Large Telescope, Subaru, and Gemini), only the disk regions beyond a few au can be spatially resolved. \rev{The angular resolution that is actually achieved by the current observations depends on the wavelength regime and, in the showcase examples of Figure\,\ref{fig:gallery}, ranges from 25\,mas to 70\,mas. Therefore, for objects in Taurus and Lupus (e.g., GG Tau and IM Lup), this translates to scales of 4 to 10\,au. In the more distant star forming region of Orion (e.g. GW Ori), we are still able to resolve scales between 10 and 30\,au. To give an impression of the different extents of the systems and of the observed sub-structures, each image in Figure\,\ref{fig:gallery} is accompanied by a scale bar of 50\,au in the lower left corner.}

To reach the theoretical diffraction limit with a large ground-based telescope, the disrupting effect of the atmosphere needs to be compensated for. This requires the employment of active or adaptive optics systems (\citealt{Babcock1953}). This technology in particular has seen crucial advancements in the past decade, leading to the latest generation of so called "extreme" AO systems (\citealt{Fusco2006,Macintosh2006,Jovanovic2015,Guyon2018}). \rev{Nowadays, images of planet-forming disks are routinely obtained at wavelengths ranging from I-band ($\sim0.8\,\mu m$) to K-band ($\sim2.2\,\mu m$), as is clear from Figure\,\ref{fig:gallery}. The observations at different wavelengths were initially driven by technical considerations. K-band and H-band observations are generally favored due to the increase stability of the atmosphere at longer wavelengths. Optical observations deliver superior angular resolution, but can only be conducted for very bright target stars due to adaptive optics requirements.}

Another requirement is the need to achieve high contrast between the primary star and the scattered light from circumstellar dust in the disk, \rev{as scattered light from the disk is significantly fainter than the star}. For gas-rich flared disks, \rev{the observed average} contrast \rev{between the polarized and stellar flux} is on the order of 10$^{-2}$ to 10$^{-4}$ (\citealt{Garufi2018}, see Sect.\,\ref{sec:brightness}).
Blocking star light first requires the use of a coronagraphic mask, but is generally optimized by combining coronagrapic hardware design with advanced observational and post-processing techniques, as discussed below. The scientific progress in the field has been enabled by significant advancements in all three areas. 


\subsection{Differential imaging techniques}
\label{sec:diff}
Even with the use of AO and state-of-the-art coronagraphs, additional image post-processing is needed to disentangle the stellar speckle halo from scattered light originating in the circumstellar environment. For this purpose, differential imaging techniques are used, combining optimal observational strategies and data reduction techniques. \\

\subsubsection{Reference Differential Imaging.} The most straightforward technique is reference differential imaging (RDI). 
It relies on observations of a reference star with similar spectral properties and apparent magnitude to the scientific target, which are used to calibrate the instrument point spread function (PSF). In an idealized scenario, the instrument and sky conditions remain identical between the two observations, and thus the PSF will not change.  One can then use the reference star image to subtract the stellar light from the science target image, leaving only the disk or planet signal. \rev{This technique has the great advantage that it in principle does not reduce the throughput of the scientifically interesting signal from circumstellar disks and planets. In practice,  it} has been mainly used in space-based observations \rev{\citep{Grady1999, Weinberger1999, Grady2000, Grady2001, Debes2017, Choquet2016}}, as the PSF stability is superior relative to ground-based observations. Libraries of reference stars can be constructed using all existing observations in the same instrument mode. This strategy, pioneered for the Hubble Space Telescope, was applied to HST/NICMOS data in the Archival Legacy Investigations of Circumstellar Environments (ALICE, \citealt{Choquet2014}). In addition to choosing the optimal reference star, principal component analysis (PCA) based reduction algorithms are used to deconstruct the reference star images and fit them to the science star.  When a bright circumstellar disk is present, this can result in over-fitting the data images. \rev{In practice, this over-fitting significantly limits the throughput of circumstellar signal, especially for disks that appear bright in scattered light.
Over-fitting can be counteracted by iterative approaches that disentangle disk and stellar signal. One recent advancement was made by \cite{Ren2018}, who use non-negative matrix factorization for this purpose.}\\
With the advent of extreme AO systems leading to better correction of the atmospheric perturbations,  RDI is now maturing as a differential imaging technique \citep{Boccaletti2021}. Recently, \cite{Wahhaj2021} presented results obtained with VLT/SPHERE on the HR8799 system, obtaining a \rev{point-source} contrast limit of 11.2 magnitudes at 0.1\arcsec{}, through alternated observations of science target and reference star. \rev{This strategy is now increasingly used in ongoing programs.} 
However, finding the optimal reference star can be difficult, especially in regions of the sky offset from the galactic plane where the stellar density is lower.  Since changing atmospheric conditions usually limit the contrast achieved with RDI \rev{from the ground}, other techniques rely on the science data itself to disentangle stellar and circumstellar light.\\

 \begin{figure*}[!t]
 \centering
  \includegraphics[width=0.9
  \textwidth]{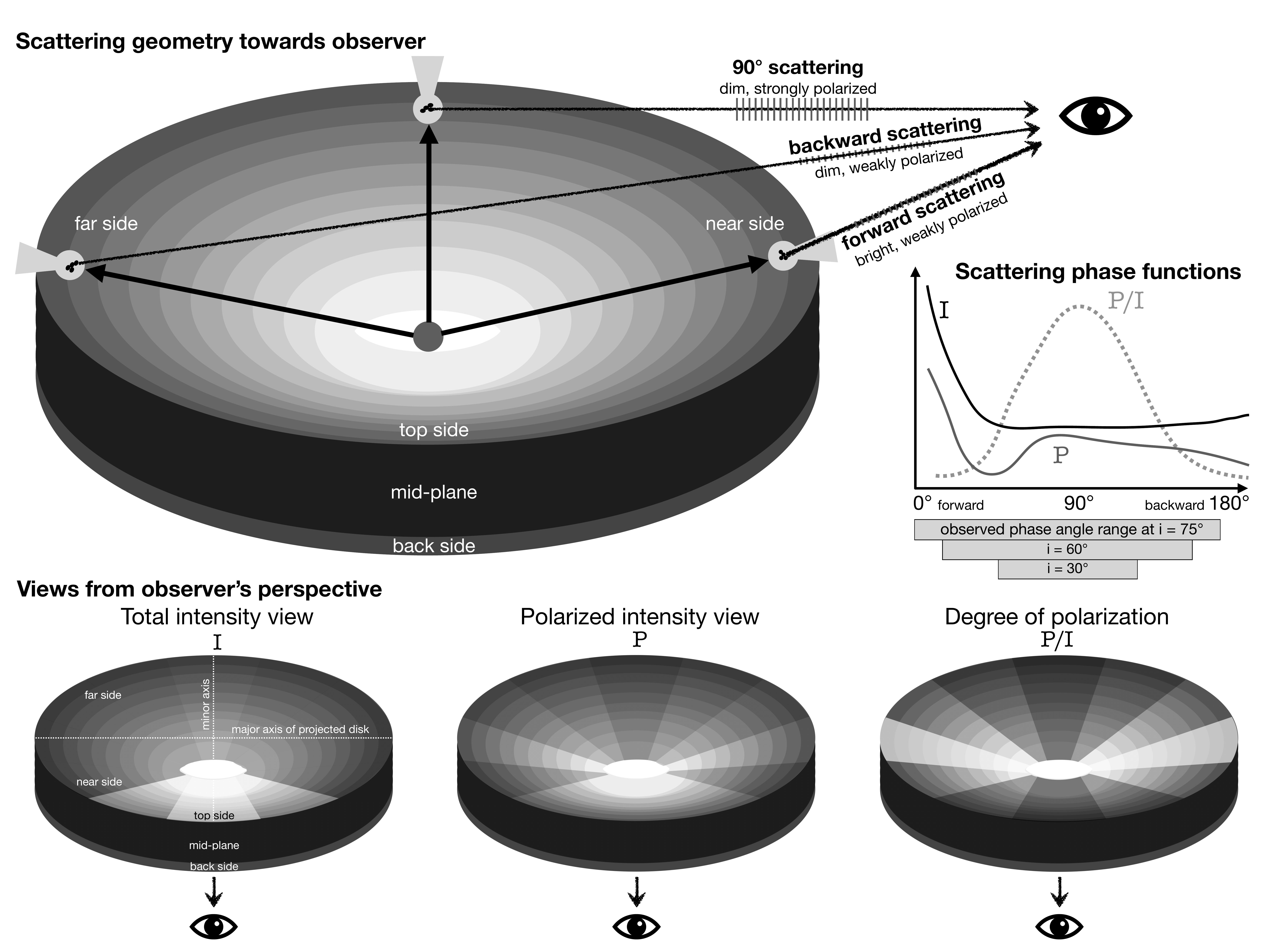}
  \caption{\small Schematic representation of the scattering geometry, viewing geometry and the influence of phase functions.  The \textbf{top panel} shows the scattering locations in which forward ($\approx 0\degree$), backward ($\approx 180\degree$), and $\approx 90\degree$ scattering take place for a viewer that is off to the right-hand side.  For the case of \rev{grains/aggregates} that are somewhat \textit{larger than the wavelengths of observation} (which seems to be typical for the studied disks), the key-hole symbols highlight the significant forward scattering enhancement at each scattering point. The total intensity $I$ and polarized intensity $P$ produced by the scattering event are sketched in \textbf{the plot}, showing the phase functions as a function of scattering angle from 0\degree (forward scattering) to 180\degree (backward scattering). Below the plot, we indicate the scattering angles that actually occur in a disk seen under specific inclinations, assuming a height of the scattering surface in the outer disk of $0.15r$.  Note that the smallest scattering angle observable will actually be limited by the inner working angle of the observation. The \textbf{bottom row} shows the brightness variations of the disk surface one would expect just from the properties of the sketched phase functions, this time using the perspective of the reader, clearly showing the effects of enhanced forward scattering by large \rev{grains/aggregates}.
  In contrast, in a disk in which \textit{small (sub-micron) grains} are dominant, the intensity phase function will be nearly constant (isotropic scattering), and the intensity view would present a much more evenly bright disk surface. The polarized intensity would peak along the major axis of the projected disk (90\degree scattering), resembling the degree of polarization view shown at the bottom right.}
  \label{fig:scatgeometry}
  \end{figure*}

\subsubsection{Angular Differential Imaging.} 
The most widely used technique is angular differential imaging (ADI). ADI was originally designed for the detection of faint thermal emission from wide-separation extrasolar planets (\citealt{Marois2006}).
The ADI technique uses the apparent rotation of the stellar field on the sky during a night, when observed with an altitude-azimuth mounted telescope. 
Instruments on such telescopes usually have a K-mirror build in, the so-called "de-rotator", to field-stabilize the image of the sky on the detector. 
For ADI observations, field-stabilization is switched off, and instead all optical components relative to the telescope entrance pupil are stabilized. In this so-called "pupil-stabilized" mode, one attempts to keep the instrumental PSF and the unresolved stellar image stable, while off-axis stars, planets or disk signal will show apparent rotation, enabling to clearly identify those. 
This technique in a simpler form was pioneered with the Hubble Space Telescope ("roll-subtraction") where the entire telescope was rolled around the optical axis between exposures (\citealt{Schneider2003}). Different ADI approaches to deconstruct and subtract the stellar light are available, for example, the principal component based analysis \citep[KLIP, PynPoint packages;][]{Soummer2012, Amara2012}, or locally optimized combination of images \citep[LOCI package;][]{Lafreniere2007}. 

ADI has the advantage compared to RDI that the stellar light contribution is determined from the science star data itself. However this comes with the disadvantage of signal suppression, often referred to as "self-subtraction".
This is caused by the temporal overlap of the planet or disk signal with itself on the detector, and depends on the amount of field rotation that is achieved during the observations.  The field rotation is given by the change of the parallactic angle and is  a function of the observing time, the location of the observatory, the declination of the star and the local sideral time. Long observations are needed, but in practice, the instrumental drift of the PSF and changing weather conditions are limiting factors \citep{Cantalloube2019}. 

For point sources such as planets, this signal suppression can be well characterized by injection of model point sources into the data before reduction.
This is however challenging in the case of extended disks with an unknown brightness distribution. A typical result is a strong signal suppression along the minor axis of the disk, leading to broken ring structures \rev{\citep{Milli2012,Perrot2016,Ginski2016,deBoer2016}}.  This effect also complicates the detection of embedded planets in disks. The presence of extended disk signal leads to a higher noise floor in the ADI processed data. Extended disk signal can also lead to false-positive planet signals when confronted with aggressive post-processing.  \cite{rameau2017} and \cite{follette2017} demonstrated this effect for the prominent case of the HD100546 system for which multiple planet candidates were suggested in the past (\citealt{Quanz2013a, Brittain2013, Currie2015}). Signal suppression is a function of separation from the center of rotation as well as the inclination of the disk. The smaller the separation from the center and the smaller the inclination of the disk, the stronger is the signal suppression.
Therefore, ADI is particularly ill suited to detect disks in face-on viewing geometry or small disks. \\


\subsubsection{Polarization Differential Imaging.} 
\rev{The use of polarization differential imaging (PDI, \citealt{Kuhn2001}) successfully enabled the imaging of circumstellar disks.} When stellar light is scattered off by dust grains, the resulting light is partially linearly polarized. The degree of linear polarization depends on the scattering angle \rev{between $0 \degree$ (forward scattering) and $180\degree$ (backward scattering)\footnote{\rev{Note that in solar system research, the scattering angle is sometimes defined such that that 0\degree corresponds to backward scattering, not forward scattering.  We use the convention that is commonly used in the context of light scattering theory and astronomical observations.}}, and} is typically peaking at 90$^\circ$, i.e., close to the ansae of the disk. Figure~\ref{fig:scatgeometry} provides a schematic representation of stellar light scattered by a disk. 

The latest generation of instruments include a polarizing beam-splitter or a non-polarizing beam-splitter combined with polarization filters (\citealt{Tamura2006,Perrin2015,Schmid2018,deBoer2020}). These are usually combined with a half-wave plate that modulates the polarization direction of the incoming light and allows to split it into two orthogonal, linearly polarized components, which are imaged simultaneously on the detector. Since the star exhibits only a low level of polarization, the stellar light will be nearly identical in both beams, while the polarized scattered light from the disk will differ. As the central star illuminates all directions of the surrounding disk, the disk will exhibit an azimuthal orientation of the electric field vector for single scattering events. This leads to a so-called "butterfly" pattern when one subtracts the two orthogonal polarized beams from each other, i.e., one polarization direction will show as two positive lobes in the image and the other as perpendicular aligned negative lobes.
The strength of the PDI technique, is that as orthogonal polarized beams are recorded simultaneously or nearly simultaneously (depending on the instrumental design), there is no temporal variation of the instrument PSF. In addition, there is no signal self-subtraction as in ADI, due to the different polarization directions in each part of the disk.

The PDI technique follows the Stokes formalism (\citealt{Stokes1851}), in which the polarization state is fully described by the Stokes vector~$S$: 

\begin{equation}
    S = \begin{bmatrix} I \\ Q \\ U \\ V \end{bmatrix} 
\end{equation}

where $I$ is the total intensity of the light, $Q$ the vertical and horizontally linearly polarized light, $U$ the linearly polarized light rotated with 45$^\circ$ with respect to $Q$, and $V$ contains the circularly polarized light. The latter is typically not measured. 
\rev{$Q$ and $U$ images are obtained by rotating the astrophysical polarized signal into the frame of the polarizer or polarizing beam-splitter within the instrument. One major consideration when attempting to record any polarized signal is that reflections within the instrument will lead to additional instrumental polarization (see \citealt{Snik2013} for a detailed discussion). These can be partially cancelled by the so-called "double-difference' method (\citealt{Bagnulo2009}) that records not only $Q$ and $U$ frames but $Q^+$, $Q^-$, $U^+$ and $U^-$. The difference between $Q^+$ and $Q^-$ ($U^+$ and $U^-$) is that the HWP is rotated by 45$^\circ$, which flips the sign of the incoming polarized signal, while instrumental polarization introduced downstream of the HWP retains its original sign. If the HWP is placed early in the path of light within the instrument, then a subtraction of $Q^+$ and $Q^-$ ($U^+$ and $U^-$) will cancel the majority of instrumental polarization, while the astrophysical signal is retained. However even using this method, some instrumental polarization, upstream of the HWP is retained. Furthermore, the stellar light itself may exhibit low levels of polarization, due to interstellar dust or local dust. Both of these will significantly affect the contrast achievable with PDI and need to be removed. To do so, it is assumed that instrumental and stellar linear polarization will to first order be proportional to the total intensity of the stellar light. Thus the usual approach is to measure the total intensity and polarized intensity signal in an area of the image for which no disk signal is expected, or for which it is expected that the stellar and instrumental polarization signal will be dominant. One can then derive the proportionality factor between total and polarized intensity and subtract a thus scaled version of the total intensity image from the double difference corrected Stokes $Q$ and $U$ images. This removes simultaneously the remaining instrumental polarization and the stellar polarization. This critical technique was pioneered by \cite{Canovas2011} and \cite{Hashimoto2012} and is now a standard component of PDI image processing.\\     }
To obtain a polarized intensity image of the disk, one can combine the Stokes Q and U components $P\!I = \sqrt{Q^2+U^2}$. However, as this involves squaring the $Q$ and $U$ images, the $P\!I$ images often suffers form a faint halo in the areas most strongly affected by photon noise. It is therefore more convenient to instead use the azimuthal Stokes parameters (\citealt{Monnier2019, deBoer2020}):

\begin{equation}
\begin{split}
 Q_\phi = -\,Q \cos(2\phi) - U \sin(2\phi)\\
 U_\phi = +\,Q \sin(2\phi) - U \cos(2\phi)   
\end{split}
\end{equation}

where $\phi$ is the azimuth angle in the image plane.
Using this formalism all azimuthally polarized signal is positive in $Q_\phi$ and radially polarized signal is negative in $Q_\phi$. $U_\phi$ contains signal that is 45$^\circ$ offset from either polarization direction. In practice $Q_\phi$ and $P\!I$ are nearly identical, with the more favorable noise properties in the $Q_\phi$ image. If the polarized light is due to single scattering events only, $U_\phi$ contains no signal and can be used as a convenient noise estimator. However, this breaks down for disks seen under high inclination, where multiple scattering along the line of sight becomes a significant factor (\citealt{Canovas2015}).   

PDI as a technique is now matured to the point that it can be used for the characterization of the innermost disk regions and directly imaged extrasolar planets. Asymmetric dust distributions in regions smaller than the telescope resolution element (e.g., inner disk, CPDs, planetary clouds)  will introduce a residual polarization. These effects are generally small, on the order of 0.1-1\% (\citealt{Stolker2017a}), but can become extreme, for edge-on viewing geometries as in the case of CS\,Cha\,b, which yielded a degree of linear polarization of 13.7$\pm$0.4\% in the J-band \citep{Ginski2018}. 
Similarly, \cite{vanHolstein2021} used unresolved polarimetric measurements of the sub-stellar companion in the DH\,Tau system to show that it must be surrounded by an accretion disk. These examples demonstrate the achieved maturity of the PDI technique, as well as the important insights that can be gained from the polarization state of unresolved sources.\\

\subsubsection{Spectral Differential Imaging.}  
A way to overcome temporal PSF variations is offered by spectral differential imaging (SDI). High contrast imaging instruments have integral field spectrographs (IFS) with low spectral resolution \citep[R $<$ 100;][]{Claudi2008,Larkin2014,Groff2015}, providing images at different wavelengths. SDI is based on the fact that the stellar speckle halo scales radially with wavelength, while any \rev{continuum source} remains nearly identical with wavelength. The stellar speckles can therefore be \rev{subtracted} after rescaling the images, revealing the disk or planet signal. SDI has the advantage that images at all wavelengths are taken simultaneously, such that temporal PSF changes or changing weather conditions are not problematic anymore. However, this technique assumes that the instrument response is identical across wavelengths. This can be a problem in particular for ground based AO imagers, as the correction of the wavefront aberrations is wavelength-dependent with better correction obtained at longer wavelengths. In addition, the radial shift applied when re-scaling the images depends on the wavelength range covered by the instrument. Finally, SDI suffers from similar signal suppression effects as ADI imaging in the radial direction and is thus not ideal for the imaging of extended structures.  At higher spectral resolution R$>$1000, specific spectral lines can be resolved, and atmospheric or accretion tracers of young planets can be observed. This technique was used to detect the second protoplanet in the PDS\,70 system, discussed in section~\ref{sec:PDS70}.\\

\subsection{Facilities and sample}

\begin{figure*}[t]
\centering
    \includegraphics[width=0.98\textwidth]{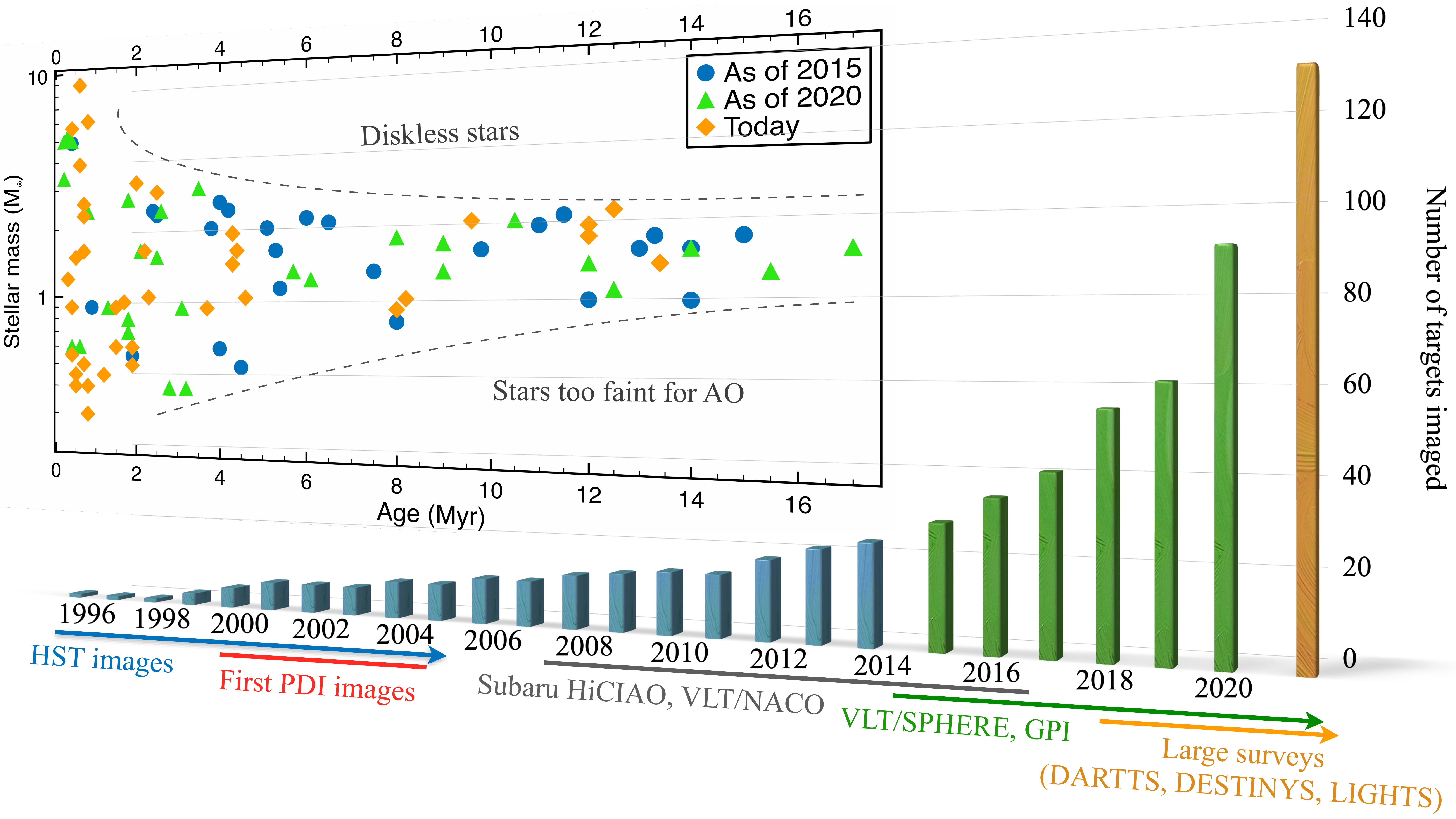}
  \caption{Total number of protoplanetary disks imaged per year. \rev{The count includes targets with observations (both detections and non-detections) that were published for the first time.} The inset diagram shows the evolving range of stellar properties probed by the available sample. The count does not include proplyds and disks seen in silhouette.}
  \label{fig:disks_imaged}
 \end{figure*}

Pioneering images of externally illuminated protoplanetary disks (proplyds) and disks seen in absorption against the bright nebular background were obtained in the Orion Nebula with the Hubble Space Telescopes \citep{Odell1993, Odell1994}. \rev{As is clear from Figure\,\ref{fig:disks_imaged},} only a few disks had been imaged at the time of \rev{PPV} \citep{Watson2007}, and most of them were edge-on disks seen in silhouette or very extended disks \citep[e.g.,][]{Grady2000, Grady2001, Krist2000, Stapelfeldt2003}. The time between \rev{PPV and PPVII} marks the transition between space- and ground-based facilities as the main carrier of optical and near-IR images of protoplanetary disks, following pioneering work by \rev{\citet{Roddier1996}}, \citet{Pantin2000}, \citet{Kuhn2001}, \rev{\citet{Itoh2002}}, and \citet{Apai2004} with the \rev{University of Hawaii adaptive optics system,} the Adaptive Optics Near IR System (ADONIS) at ESO, the United Kingdom  IR  Telescope  (UKIRT), \rev{the Subaru/CIAO,} and the VLT/NACO facilities, respectively. Nowadays, the vast majority of scattered light images \rev{at the optical and near-IR wavelengths} are acquired with 8-m, AO- and polarimeter-equipped, telescopes, in particular with VLT, Subaru, and Gemini \rev{in PDI mode (see Sect.\ref{sec:diff})}.

The first instrument to perform systematic PDI observations of disks was the Subaru/HiCIAO instrument, through the SEEDS program that observed \rev{tens of protoplanetary disks, with 18 observations so-far published}, half of which being bright T Tauri stars and the other half, Herbig Ae/Be stars \citep[e.g.,][]{Hashimoto2012, Kusakabe2012, Mayama2012, Tanii2012, Takami2013, Follette2013, Tamura2016, Mayama2020}. PDI was also successfully employed with VLT/NACO to image the disk around a handful of Herbig Ae stars \citep{Quanz2011, Quanz2012, Quanz2013a, Canovas2013, Garufi2013, Garufi2014b, Avenhaus2014b, Avenhaus2014a}. This first PDI campaign was limited by the performance of the AO system available at the time enabling detection of only bright disks (typically 10$^{-2}$ times the central star, see Sect.\,\ref{sec:brightness}) around bright stars (typically $R$ mag $<$ 10). As of 2015, the two dozens targets imaged with PDI were mainly 1--3 M$_\odot$ stars older than 4 Myr (see Figure\,\ref{fig:disks_imaged}, top), providing therefore a very biased sample.

The sample \rev{of observed targets} was then significantly expanded with the advent of the extreme-AO system driving VLT/SPHERE \citep{Beuzit2019} and GPI \citep{Macintosh2014}. Between 2015 and 2020, the sample size tripled, reaching more than a hundred. In addition, the range of stellar properties probed by these images stretched to younger ages (1--2 Myr) and lower mass (0.3 M$_\odot$, see Figure\,\ref{fig:disks_imaged}, top). The stability of the systems employed and the maturity of the post-processing techniques have motivated systematic observations performed under specific observing programs such as DARTTS-S \citep[Disks ARound TTauri Stars with SPHERE,][]{Avenhaus2018, Garufi2020a}, LIGHTS \citep[Large Imaging with GPI Herbig/TTauri Survey,][Rich et al.\,in prep.]{Laws2020}, and DESTINYS \citep[Disk Evolution Study Through Imaging of Nearby Young Stars,][]{Ginski2020, Ginski2021}. 

Despite the significant progress in alleviating the sample bias, the stellar parameter space that is currently covered is still impacted by technical limitations. The minimum stellar brightness necessary to drive the AO at the telescope is in fact preventing us from accessing faint embedded young stars, the brown dwarf regime, and stars lighter than 0.3\,M$_\odot$. As is clear from Figure\,\ref{fig:disks_imaged}, the observed stars distribute in a sort of cusp with the upper desert due to absence of disks in old massive stars and the lower desert imposed by the current limits of AO guiding systems.





\section{Theoretical background \rev{of disk substructures}} \label{sec:theory}
To provide context for the observed substructures in \rev{disks} we present a brief review on how such features could be generated. Both scattered light and sub-millimeter continuum emission in these sources are probing dust grains. The origin of the observed substructures in disks is therefore intimately related to the motion of dust particles embedded in a gaseous medium. \rev{As the gas moves at sub-Keplerian velocities, due to its pressure support, and the dust grains move at Keplerian velocities, a velocity difference \rev{exists} between gas and dust grains}. Drag forces then act on dust grains and influence their motions.  Small \rev{grains} are well coupled to the gas, and their distribution is expected to be representative of the distribution of the gas. Regions where dust \rev{grains} can be lifted high above the disk midplane (due to high surface density or strong vertical mixing) will be exposed to direct stellar light that can be scattered towards the observer.  On the other hand, low \rev{densities} and weak turbulent mixing remove grains from the direct exposure to stellar light, leading to dark regions in the disk, \rev{while drag} forces will cause larger grains to move towards higher pressure regions.  
Long-term stable pressure maxima and minima (often implying surface density maxima and minima) will lead to variations in the distribution of both small and large grains, creating observable features \rev{at various tracers}. The origin of these substructures is therefore related to the mechanisms that can produce long-term stable pressure variations. In the following, we provide a brief overview of such mechanisms. \\




\subsection{Mechanisms not invoking planets} 
\rev{Radial gradients of temperature and density in a disk form condensation fronts, regions where the gas transitions from a gaseous to solid form. These condensation fronts typically contain water and carbon oxides.} At these fronts the change in chemistry and opacity can lead to a non-monotonic pressure gradient, with the possibility to collect dust particles \citep{2006Icar..181..178C}, or induce grain growth
\citep{2015ApJ...806L...7Z}.
Using a hydrodynamical two-fluid approach it was suggested that an instability arises through the vertical settling of dust in protoplanetary disks which produce toroidal gas vortices that are able to collect dust in them, leading to observable ring-shaped features \citep{2015MNRAS.453L..78L}.

The formation of organized structures in disks has been seen also in a variety of magnetohydrodynamical (MHD) simulations, both in local and in global simulations.  In this summary, we focus only on global simulations because the local simulations only yield spatial information on small scales of the order of the disk thickness (pressure scale height), while the observed features cover a large spatial extent. For more details, we refer the reader to the Chapter by \textit{Lesur et al.} In non-ideal simulations that included a transition between a magnetically active  region where turbulence is driven by the
magneto-rotational instability (MRI) and a dead-zone, where non-ideal effects quench the \rev{MRI,} \citet{2016A&A...590A..17R} observed dust collection near the interface between the two regimes due to a presence of a pressure maximum. \citet{2019A&A...625A.108R} provide an overview of MHD simulations that show ring occurrence. Generally, configurations with net magnetic field can show ring-like features, even in the ideal-MHD case, where
the rings represent themselves as a succession of sub-Keplerian and super-Keplerian bands \citep{2017A&A...600A..75B}.
In disks that are unstable to the MRI, 
\citet{2016A&A...589A..87B} found self-organized toroidal structures in unstratified three-dimensional disks when including the Hall-effect in the MHD-equations. 
In global axisymmetric simulations using ambipolar diffusion \citet{2020A&A...639A..95R} show that the zonal flows (rings) are closely 
linked to the existence of a MHD wind. In their simulations, the ring width and separation is roughly compatible with observed features. 

An important non-axisymmetric feature in disks are vortices, i.e., localized regions which, in a frame moving with the mean flow, show a reversed (anti-cyclonic) rotation. Such vortices are regions of increased density and pressure and small embedded dust particles will collect in them leading to observable, crescent-shaped features. In hydrodynamical disks without an embedded planet, two scenarios can lead to such vortices. The first is the subcritical baroclinic instability \citep[SBI,][]{2003ApJ...582..869K,2010A&A...513A..60L}, which requires a negative radial entropy gradient in the disk and finite amplitude perturbations. The second is the vertical shear instability \citep[VSI,][]{2013MNRAS.435.2610N,Barraza2021,Blanco2021} suggested to generate turbulence in disks which requires a vertical gradient in the angular velocity and short cooling times. In thicker, nearly isothermal disks large scale vortices can be generated as shown in 3D hydrodynamical simulations \citep{2016MNRAS.456.3571R,2020MNRAS.499.1841M}. 

\subsection{Gap creation by planet-disk interactions} 
\label{theory-gaps}
Planet-disk interactions can also trigger a variety of features that can be observed. This process is based on the effect that a planet created a wake that travels through the disk as a spiral wave.  

Sufficiently massive planets embedded in a disk deplete the gas at the orbital location if the thermal and viscous criteria for gap-opening are met. Such an annular gap is caused by angular momentum transfer from the planet to the ambient gas induced by the spiral density arms that are created by the planet, as showed in full hydrodynamical simulations  \citep{1999ApJ...514..344B,1999MNRAS.303..696K}. The gap depth and width depend on the disk viscosity 
\citep{2006Icar..181..587C}. Above a critical mass of few ten $M_{\mathrm{Earth}}$ (depending on the disk viscosity and temperature), density (pressure) maxima inside and outside the gaps exist. Dust can collect there and lead to the observed ring-like features where the dust density \rev{divergence} exceeds that of the gas \citep{2006A&A...453.1129P}. 

A striking application of these ideas is the interpretation that possibly all the \rev{multiple} rings observed in sub-millimeter observations are created by embedded planets \citep{Zhang2018, Lodato2019}.  We refer to the Chapter by \textit{Bae et al.} for a comparison with the exoplanet population. For several cases it was noted that one single planet is able to produce a multitude of rings \citep{2017ApJ...843..127D}. However, this effect turned out to be a consequence of the over-simplified thermodynamics used to describe the disk temperature in the simulations (often a power law for the temperature radial profile). 
The reason for the appearance of multiple rings was traced back to the locally isothermal approximation which results in an incorrect angular momentum transport in disks \citep{2019ApJ...878L...9M}.
\citet{2020MNRAS.493.2287Z} show that for a cooling time that equals the dynamical timescale, $\tau_{\mathrm{cool}} \sim \Omega^{-1}$, the strength of the spiral arms is weakest, while for shorter (more isothermal) or longer (more adiabatic) cooling the strengths of the spirals become stronger again. As a consequence the number of rings created by a single planet varies with the cooling timescale \citep{Facchini2020}. \citet{2020A&A...637A..50Z} showed in fully radiative simulations which included viscous and stellar heating that a multi-ring system \rev{such as the one observed in AS209 with ALMA \citep{Guzman2018}} is better explained by multiple planets instead of a single one, emphasizing the need to consider  radiative effects. If planetary migration is included, \citet{2020MNRAS.493.5892W} showed that a single planet with different migrations speeds
can in principle generate multiple rings in the dust distribution which are, however, slowly dissipating with time via disk turbulence. For more details on migration, we refer to the Chapter by \textit{Paardekooper et al.} 

\subsection{Spiral waves}
\label{theory-spirals}
Spiral arms can be produced in disks, and such spiral arms can differ in their pitch angle $\beta$, which is the angle that the spiral makes with a circle around the star. First, a planet embedded in the disk will lead to wakes in the form of spiral arms that are sheared out by the Keplerian motion. The value of $\beta$ is determined by the local temperature (sound speed) in the disk. Using linear theory, valid for planets lighter than the thermal mass, $M_{\rm th} \sim h^3 M_*$, where $h$ denotes the disk aspect ratio, the shape of the spirals was given for disks with constant $h$ by \citet{2002MNRAS.330..950O,Muto2012}.  
Deviations occur for massive planets where the spirals turn into strong shocks \citep{2015ApJ...813...88Z} and when the heating by the spirals is included \citep{2020A&A...633A..29Z}. In these cases the pitch angle becomes larger.  A second possibility would be a massive (stellar) companion object, 
that triggers the formation of tidal arms in the inner disk which appear as wide spirals with a large pitch angle  \citep{2015ApJ...809L...5D,2016ApJ...826...75D}. In \rev{a} circumbinary disk,  the gravitational torque exerted by the binary leads to the formation of a large eccentric inner cavity in the circumbinary disk with highly reduced mass and large asymmetric features can be produced in the disk such as blobs, eccentricities and spiral arms \citep[see for example][]{2017A&A...604A.102T,2017MNRAS.464.1449R,2020A&A...639A..62K}. Another, third possibility, is the formation of spiral features in more massive disks where the disk's own self-gravity cannot be neglected, usually for disk masses of about $\sim 1/10$ of the stellar mass. 
In simulations that include disk self-gravity and a simplified cooling presciption it was shown using, Smoothed Particle Hydrodynamics (SPH) simulations that the disks settle to an equilibrium state where the induced turbulent heating by the spiral arms match the cooling rate where the Toomre parameter is unity $Q \sim 1$ throughout the disk \citep{2009MNRAS.393.1157C}. \rev{Additionally, in recent 3D grid simulations it was shown  that} the occurrence of the spiral arms is an intermittent process and the spirals do not show long-term global coherence \citep{2021A&A...650A..49B}. Both simulations indicate that the pitch angles are confined to a narrow range $\beta \sim 15^\circ$, relatively independent of the stellar mass and disk radius.

Planets can also induce non-axisymmetric features. As \rev{an annular} gap is created by a massive planet, steep radial gradients of the surface density, pressure and a change in the rotational profile occur. This can trigger a Rossby-wave instability \citep[RWI,][]{1999ApJ...513..805L} near the gap edge creating a vortex \citep{2005ApJ...624.1003L}. Numerical studies have shown that vortex properties depend on various physical processes such as turbulent viscosity and disk self-gravity. Lower viscosity allows vortices to live longer \citep{2006MNRAS.370..529D}, whereas the inclusion of self-gravity tends to weaken vortices, shortening their lifespan \citep{2017MNRAS.471.2204R}.

Most theoretical studies consider planets orbiting in the plane of the disk. 
However, the presence of a massive \rev{planetary or sub-stellar} companion on an inclined orbit can induce a warp in the disk \citep{xg2013}, and relative misalignment of specific disk radii. The condition for the tilting to occur is that the companion  angular momentum is greater than the inner disk's \citep{Matsakos2017}. This is confirmed by 3D hydrodynamical simulations that show that in the case of an inclined planet, massive enough to carve \rev{an annular} gap, the inner and outer disk can be misaligned \citep{Nealon2018}, and in some cases, the disk can have a higher inclination than the planet \citep{Bitsch2013}. If the mass ratio is high, in the extreme case of an equal mass stellar binary, the inner disk can  break and freely precess \citep{Facchini2013}, leading to a variety of misalignment angles that could induce narrow or broad shadows on the outer disk \citep{facchini2018}. Secular precession resonances in the case of high stellar-companion mass ratio can also lead to significant disk misalignments  \citep{Owen2017}. Finally, the Kozai-Lidov effect can bring a massive planet located within the cavity to a high inclination in the case of a multiple systems, like HD100453, inducing the misalignment of the inner disk \citep{Martin2016,Gonzalez2020,Nealon2020,ballabio2021}.





\section{Global structure of protoplanetary disks}
\label{sec:globalstructure}

In the following section, we review observational results that enabled to constrain the global disk structures, through their spatial extent in the IR scattered light and the apparent shape of the scattering surface. We also discuss how the dust grain properties directly affect the disk appearance. 


\begin{figure}[t]
\centering
    \includegraphics[width=0.485\textwidth]{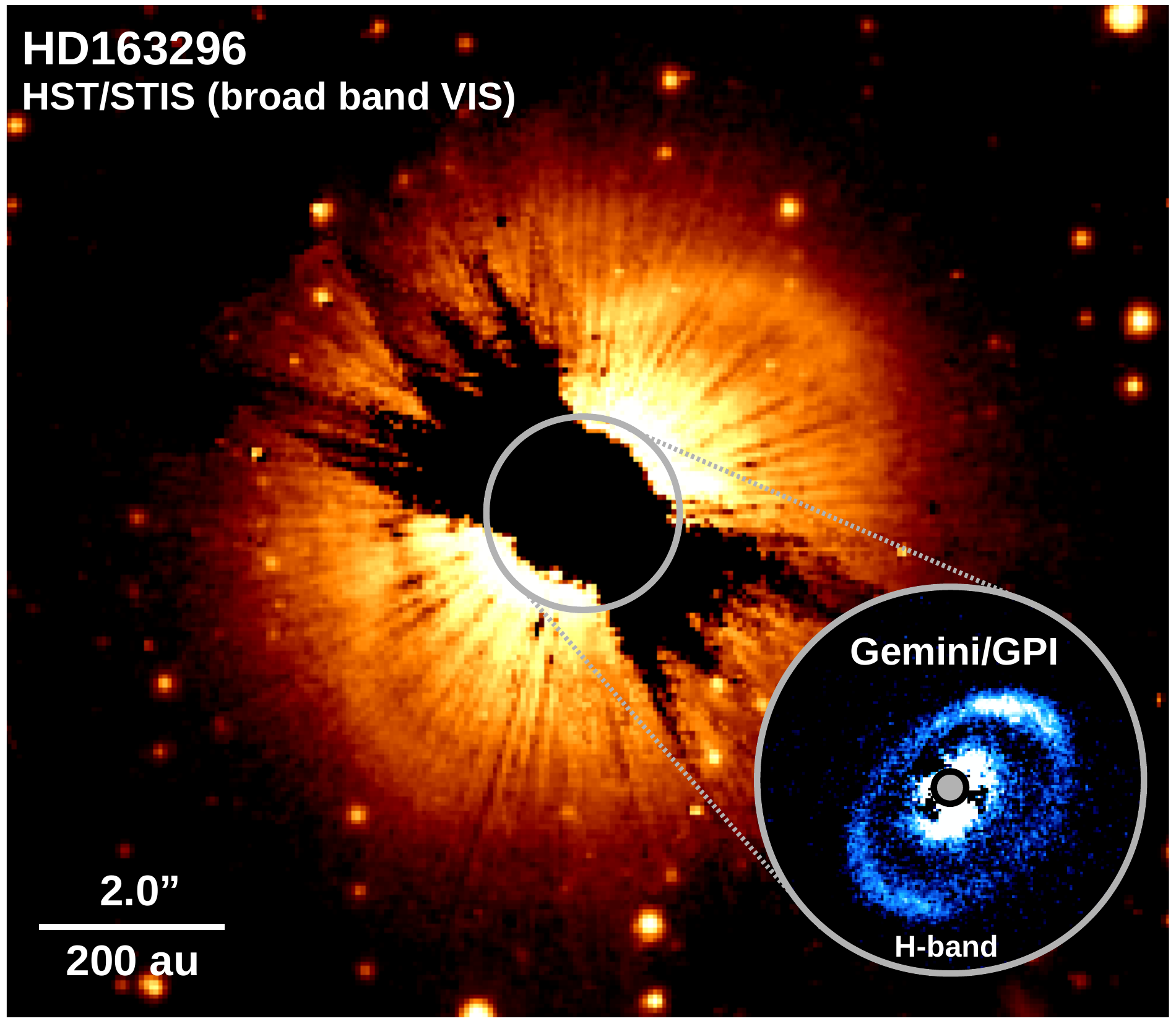}
  \caption{\rev{Hubble Space Telescope total intensity image of the disk surrounding the Herbig star HD\,163296 taken at visible wavelength range with the Space Telescope Imaging Spectrograph (STIS). The masked center area was covered by a coronagraphic mask. We show a combination of all observation epochs presented in \cite{Rich2020}. Part of this data was first presented in \cite{Grady2000}. The circular inset shows Gemini/GPI polarized light data taken in the near-IR from \cite{Monnier2017}. While lacking the same inner working angle as modern ground based facilities, the HST data is sensitive to faint extended structures typically not detected from the ground.}}
  \label{fig:HD163296-HST}
 \end{figure}

\subsection{Disk radial extent and brightness} \label{sec:brightness}
In the following, we refer to 'disk size' as being the extent of the disk in scattered light. The first targets imaged in scattered light were small-scale reflection nebulae like T Tau \citep{Nakajima1995} and exceptionally bright and extended disks like AB Aurigae \citep{Roddier1996, Weinberger1999, Grady1999}. Before 2010, only very large disks ($>$100 au in size) such as HD163296 \rev{(see Figure \ref{fig:HD163296-HST})} or SU Aur were observed with HST \citep{Grady2000, Grady2001, Grady2009, Krist2000, Pinte2008}, or as HD100546, with the first AO-systems \citep{Pantin2000, Kuhn2001, Fukagawa2006, Riaud2006}.  AO-assisted observations from 8-m telescopes then detected a dozens disks of intermediate sizes \citep[15$-$100 au in size, e.g.,][]{Mayama2012, Thalmann2015, Rapson2015}, some that were previously undetected in HST surveys \citep{Grady2005}. The successful observations of disks of intermediate sizes is due to the fact that their extent corresponds to separations of 0.1\arcsec$-$0.7\arcsec, where the stellar illumination is still significant and both the angular resolution and sensitivity achieved by current ground-based telescopes are optimal to probe the disk. On the other hand, the common employment of a coronagraph for this type of observation prevents the access to separations (and therefore disk extent) smaller than 0.1\arcsec --0.15\arcsec. The first attempt to characterize small disks ($<$15 au in extent) has recently been made with high-precision polarimetry, as mentioned in Section\,\ref{sec:diff}. The polarization state of unresolved light can be measured if the instrumental polarization and cross-talk are corrected with sufficient accuracy \citep{vanholstein2020}. The determination of the degree and angle of linear polarization from unresolved emission together with radiative transfer modeling allows one to infer the presence, orientation, and potentially inclination of unresolved disks around the primary star \citep{Keppler2018,Garufi2020a} or around sub-stellar companions \citep{Ginski2018, vanHolstein2021}. 

\begin{figure*}[t]
\centering
  \includegraphics[width=1.0\textwidth]{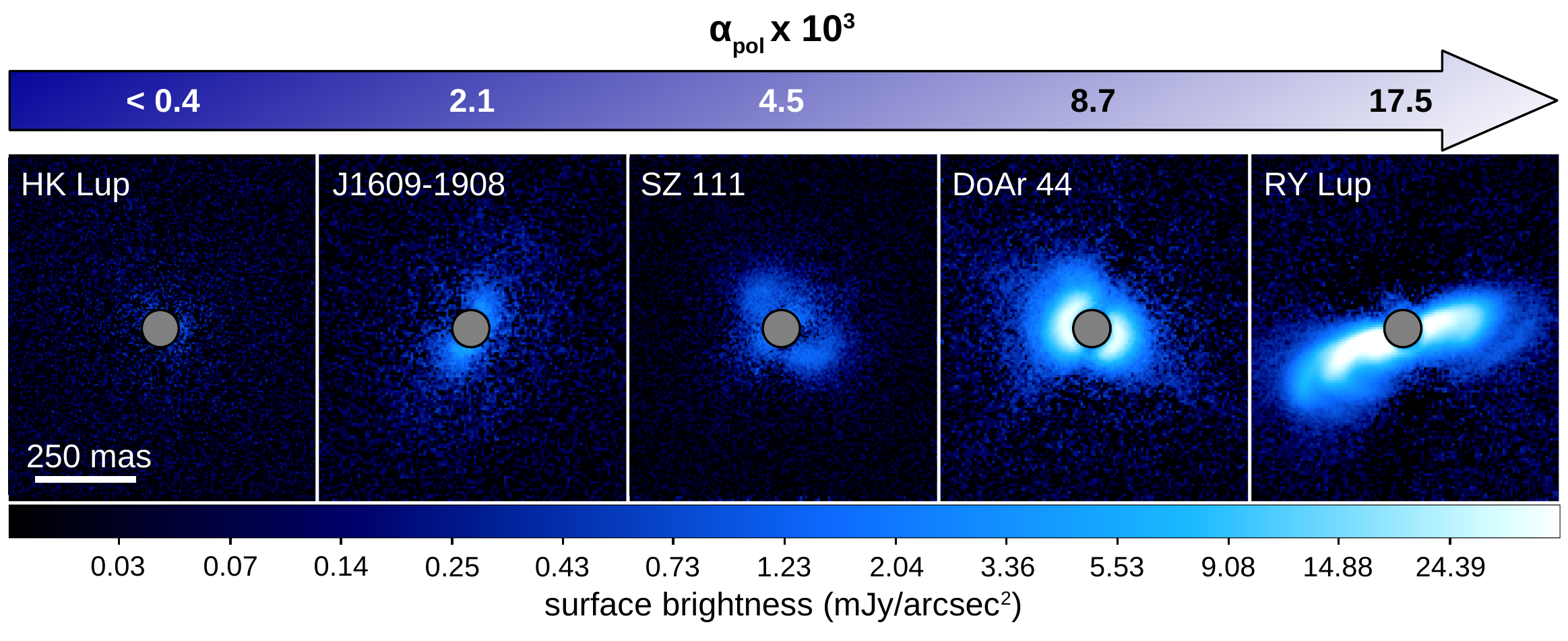}
  \caption{\small Disks observed with VLT/SPHERE sorted by polarized-to-stellar light contrast $\alpha_{\rm pol}$ (see Eq.\,\ref{Formula_albedo}) with increasing disk brightness from left to right. All images are using the same absolute color map. \rev{References are listed in Appendix~\ref{App: ref: faint gallery}.}}
  \label{fig:brightness}
 \end{figure*}

While the diversity of disk sizes can be appreciated from near-IR imaging, the determination of the disk extent solely based on a sub-set of images is questionable. An illustrative case in this regard is HD163296 \rev{(Figure \ref{fig:HD163296-HST})}, where the disk is detected out to more than 500 au from HST images \citep{Wisniewski2008} \rev{and to 440 au from Subaru/CIAO images \citep{Fukagawa2010}} but to less than 100 au from SPHERE \rev{and GPI \citep{MuroArena2018, Monnier2017}}. \rev{A similar difference between HST and SPHERE images was reported for 2MASS J1609-1908 and 2MASS J1614-1906 \citep{Garufi2020a,Walker2021}.} This discrepancy is due to the modest sensitivity of polarimetric observations from the ground as the polarized light is only a fraction of the total scattered light (see Sect.\,\ref{sec:grains}). In general, the disk outer radius with detectable signal is strongly related to the \rev{amount of detectable photons in polarized scattered light}. The observable amount of scattered light from a circumstellar disk is  determined by the following factors, summarized in Figure\,\ref{fig:scatgeometry}: ($i$) the number of photons that reach the disk surface, governed by the stellar luminosity and by the possible presence of disk material shadowing the outer disk regions (see Sect.\,\ref{sec:shadows}); ($ii$) the overall efficiency of dust grains in scattering photons at a specific wavelength (the dust albedo); ($iii$) the disk geometry and the scattering phase function, as these two elements determine the fraction of photons scattered in the direction of the observer \citep{Mulders2013a}; ($iv$) in case of polarimetric observations, the polarizing efficiency of scattering particles \citep[see Sect.\,\ref{sec:grains}; e.g.,][]{Murakawa2010}. 

\rev{The disk brightness in scattered light can be measured comparing the surface brightness of the disk $S_{\rm scat}$ with the stellar flux $F_*$ measured from the same set of images, which defines a brightness measure that is independent from the luminosity. The typical ratio between the $S_{\rm scat}$ integrated over the entire disk and $F_*$ is around $10^{-2}$ \citep[e.g.,][]{Fukagawa2010}. However, to measure the local capability of the disk to scatter the stellar radiation, we need to consider that the illumination from the star scales as a function of distance $r$ as $1/(4\pi r^2)$. Therefore, we define $\alpha_{\rm scat}$ as the ratio of the surface brightness relative to the measured stellar flux as:}
\begin{equation}
    \alpha_{\rm scat} (r)=S_{\rm scat}(r)\cdot \frac{4\pi r^2}{F_*}
\end{equation}
To alleviate the dependence of the observed \rev{brightness} on the disk inclination (through the scattering phase function), the $S_{\rm scat}$ can be measured along the disk major axis where all observed photons are scattered by angles close to 90$\degree$. \rev{Also,} to obtain a unique value representative of a disk, the contrast can be averaged between the innermost and outermost radii with detectable signal $r_{\rm in}$ and $r_{\rm out}$. \rev{The scattered-to-stellar light contrast $\alpha_{\rm scat}$ thus calculated is then:}
\begin{equation} \label{Formula_albedo_0}
\alpha_{\rm scat}= \frac{1}{r_{\rm out}-r_{\rm in}} \cdot \int_{r_{\rm in}}^{r_{\rm out}} S_{\rm scat}(r)\cdot \frac{4\pi r^2}{F_*} dr 
\end{equation}
If the scattered-light images are polarimetric, the polarized-to-stellar light contrast $\alpha_{\rm pol}$ from the observed polarized \rev{surface density} $S_{\rm pol}$ is therefore \rev{similarly}: 
\begin{equation} \label{Formula_albedo}
\alpha_{\rm pol}= \frac{1}{r_{\rm out}-r_{\rm in}} \cdot \int_{r_{\rm in}}^{r_{\rm out}} S_{\rm pol}(r)\cdot \frac{4\pi r^2}{F_*} dr 
\end{equation}
This measurement expresses the fraction of photons released by the star that are effectively scattered and polarized by the \rev{resolved portion of the} disk.
\citet[][2021]{Garufi2017} measured the $\alpha_{\rm pol}$ for a large number of disks finding values from few times $10^{-4}$ to few times $10^{-2}$ (the current detection limit for any disk being $\alpha_{\rm pol}\sim3\times10^{-4}$). Figure\,\ref{fig:brightness} provides the values of a few illustrative sources. We classify targets into three categories: faint disks ($\alpha_{\rm pol}<10^{-3}$ \rev{typically corresponding to} peak brightness of \rev{less than a} mJy/arcsec$^2$), moderately bright disks ($\alpha_{\rm pol}\sim10^{-3}-10^{-2}$ and peak brightness of few mJy/arcsec$^2$) and bright disks ($\alpha_{\rm pol}>10^{-2}$ and peak brightness of tens mJy/arcsec$^2$).

A clear dichotomy between bright and faint disks was already found in early scattered light observations of Herbig AeBe stars \citep{Grady2005}, that detected far-IR bright disks \citep[so-called Group I;][]{Meeus2001} while far-IR faint (Group II) yield non-detections. This was originally interpreted as tracing differences in the  geometry of the outer disk, with flared vs flat disks, respectively. However, recent studies found that Group I sources host disks with a large cavity ($>$10 au), imaged at near-IR or millimeter wavelengths \citep{Maaskant2013, Garufi2014b, Honda2015}. In general, \rev{this work shows} a clear correspondence between a high disk brightness and the presence of a dust-depleted cavity, suggesting that the scattered light brightness and the flared structure of the outer disk is related to the directly illuminated (and therefore bright) outer cavity edge of the disk. 

Intermediate and large disks ($>$15\,au) that appear faint in scattered light can be explained by (partial) self-shadowing \citep{Dullemond2002}. The observed anti-correlation between the disk brightness and the \rev{thermal} near-IR excess from the SED \citep{Garufi2022} supports a view where, in these objects, the disk inner rim at the dust sublimation front is puffed up \citep{Dullemond2001}, and leaves the outer disk in penumbra \citep{Dong2015}. A clear example is the non-detection of HK Lup in polarized scattered light \citep[see Figure\,\ref{fig:brightness}]{Garufi2020a} \rev{despite its large disk extent and brightness seen with ALMA} \citep{Ansdell2016}. The distribution of the disk brightness appears uniform across M-to-A spectral types \citep{Garufi2022}, and given the bias of the current census (see Sects.\,\ref{sec:method} and \ref{sec:statistics}), self-shadowed disks may be quite common. 


As explained above, the measurement of the disk extent in scattered light depends on many factors. In addition to those previously mentioned regarding telescope sensitivity, scattering properties and disk geometry, the intrinsic vertical distribution of small grains in the disk upper layers, controlled by the turbulent mixing, \rev{may be} another key factor. Scattered light measurements can therefore not be used to estimate the \textit{physical} disk extent, that should be measured in an optically thin gas tracer of the bulk disk density with ALMA. For more details on this topic, we refer to the Chapter by \textit{Miotello et al.}

\subsection{Shape of the disk surface} \label{sec:surface}
A disk in hydrostatic equilibrium directly exposed to the stellar radiation is expected to exhibit a flaring surface \citep[e.g.,][]{Chiang1997}. Such a geometry was proposed by \citet{Kenyon1987} based on the large \rev{thermal} IR excess of YSOs and confirmed with Hubble observations showing the disk silhouette \citep[e.g.,][]{Burrows1996, Padgett1999}. Constraining the exact shape of the disk surface is fundamental to obtain a comprehensive view of the thermal and chemical structure of the disk. Scattered-light images \rev{(along with images taken in PAH emission features, \citealt{Lagage2006})} are among the best methods to measure the surface shape as, \rev{in optically thick disks,} the observed photons are scattered by the uppermost disk layer. \rev{Observations of disk seen edge-on provide a direct measurement of the surface height \citep{Stapelfeldt1998, Stapelfeldt2014, Villenave2020}. This type of observation is typically not possible for ground-based telescopes because the star needed to drive the AO system is masked by the disk itself and is thus faint at optical or even near-IR wavelengths. However, there are a growing number of objects, observed from the ground, for which the backside of the disk is visible as forward scattering peak below the dark midplane (see Figure~\ref{fig:gallery} for some examples). 
When the disk midplane appears as a dark lane separating the observable front and rear disk faces  \citep[e.g. IM Lup and DoAr 25;][]{Avenhaus2018, Garufi2020a}, the $H(r)$ can be roughly measured from its width, and is found to be very large in the outskirt of such extended disks ($\sim$70 au at $r=300$ au). 
These measurements can be refined by combination with dedicated radiative transfer models \citep[e.g.,][]{Villenave2019}.}

\rev{
More generally the disk surface height can be inferred from the structures traced in the image, assuming that the disk is not eccentric. This is due to the fact that the typical flared and thus "bowl-shaped" surface of the disk will produce apparent offsets of disk structures along the minor axis, when viewed under an inclination. This effect is readily visible in our schematic figure~\ref{fig:scatgeometry}, where the rim of the near-side of the disk appears much closer to the central star than the rim of the far-side. This well known projection effect has been used by \cite{Silber2000} and \cite{Duchene2004} to measure the surface height of the GG\,Tau circumbinary disk, by tracing its inner rim at the edge of the central cavity. The same method can in principle be applied by tracing the outer rim of compact disks. This method becomes particularly powerful in disks with multiple rings (\citealt{deBoer2016}). In these cases the overall shape of the surface $H(r)$ can be measured as a function of the separation from the central star and its flaring parameter (tracing the evolution of $H(r)$ with $r$) can be fitted.}
However, fitting approaches can cause significant discrepancy in the measure of the scattering height. For example, in HD97048, \cite{Ginski2016} measured a height of 18.5\,au at 100\,au, with a flaring index of 1.73$\pm$0.05 while \cite{Rich2021} find a larger value of 2.48$^{+1.52}_{-0.3}$. \rev{This discrepancy is  due to the fact that the former study concentrated on the sharp-peaked, forward scattering side of the ring for the extraction of the ring offset and semi-major axis, while the latter considered also the far side. The reasoning behind the different approaches was that the disk far side is marginally radially resolved and thus some ambiguity may be introduced whether one traces the same height throughout the ring. It may well be possible that the true $H(r)$ function lies in between the reported results.} On the same target, the height of the emission of polycyclic aromatic hydrocarbons (PAH) was inferred to be $\sim$34\,au at 100\,au with a flaring index of 1.26$\pm$0.05 (\citealt{Lagage2006}). \cite{Avenhaus2018} applied this method to five bright T Tauri disks, finding flaring index values ranging between $\sim$1.1 and $\sim$1.6, and that the individual rings of the various disks could be fitted by an individual flaring index of 1.21 with $H(r=100{\rm \ au})=15.8$ au. \\
\rev{Obviously this method can only be applied to disks that show regular ring structures, and excludes those who have complex substructures \citep{Laws2020}. In these cases the height of the scattering surface may be inferred from measuring the scattering phase function of the dust particles. However a good knowledge of the disk surface is a priori needed to correct for the projection effects to measure the real scattering angles \citep{Stolker2016a}. This degeneracy can in principle be broken with detailed radiative transfer modelling as shown for example by \cite{Takami2014, Dong2016}. These simulations on the other hand depend on our knowledge of the dust scattering phase functions, which, as we discuss in Sect.\,\ref{sec:grains} is still poor.}



\cite{Rich2021} complemented this method with complementary measurement of the gas emission surfaces as observed with ALMA in the $^{12}$CO gas line in three disks. Such measurements can constrain the level of coupling between the small grains and the gas, as well as the turbulent mixing distributing small grains in the upper surface layers. They find that in two disks, the $^{12}$CO and scattering surface are similar at small radii, but that at large distance, the scattering surface is lower than $^{12}$CO emission surface, while for the third object, the two surfaces are similar. Radiative transfer modeling shows that a large gas-to-dust ratio could be responsible for the surface height differences, as well as changes in the physical conditions in the disk such as local heating due to a planet. Depending on the (molecular, dust) tracer considered the measured emission height will differ, and this is strongly dependent on temperature (for gas line as $^{12}$CO) and on dust properties (for scattered light).  

We note that these disk surface measurements are only available on bright disks. Little is known about self-shadowed disks, for which the $H(r)$ of the disk regions in the penumbra are expected to be significantly lower. Measurements of these values and an actual distribution of values across an unbiased sample of disks is currently unavailable due to the difficulty in observing these faint disks. Similarly, the scattering surface shape is \rev{typically unknown} for disks with complex morphology.  

\subsection{Dust properties} \label{sec:grains}

\begin{figure}[!t]
\centering
        \includegraphics[width=0.43\textwidth]{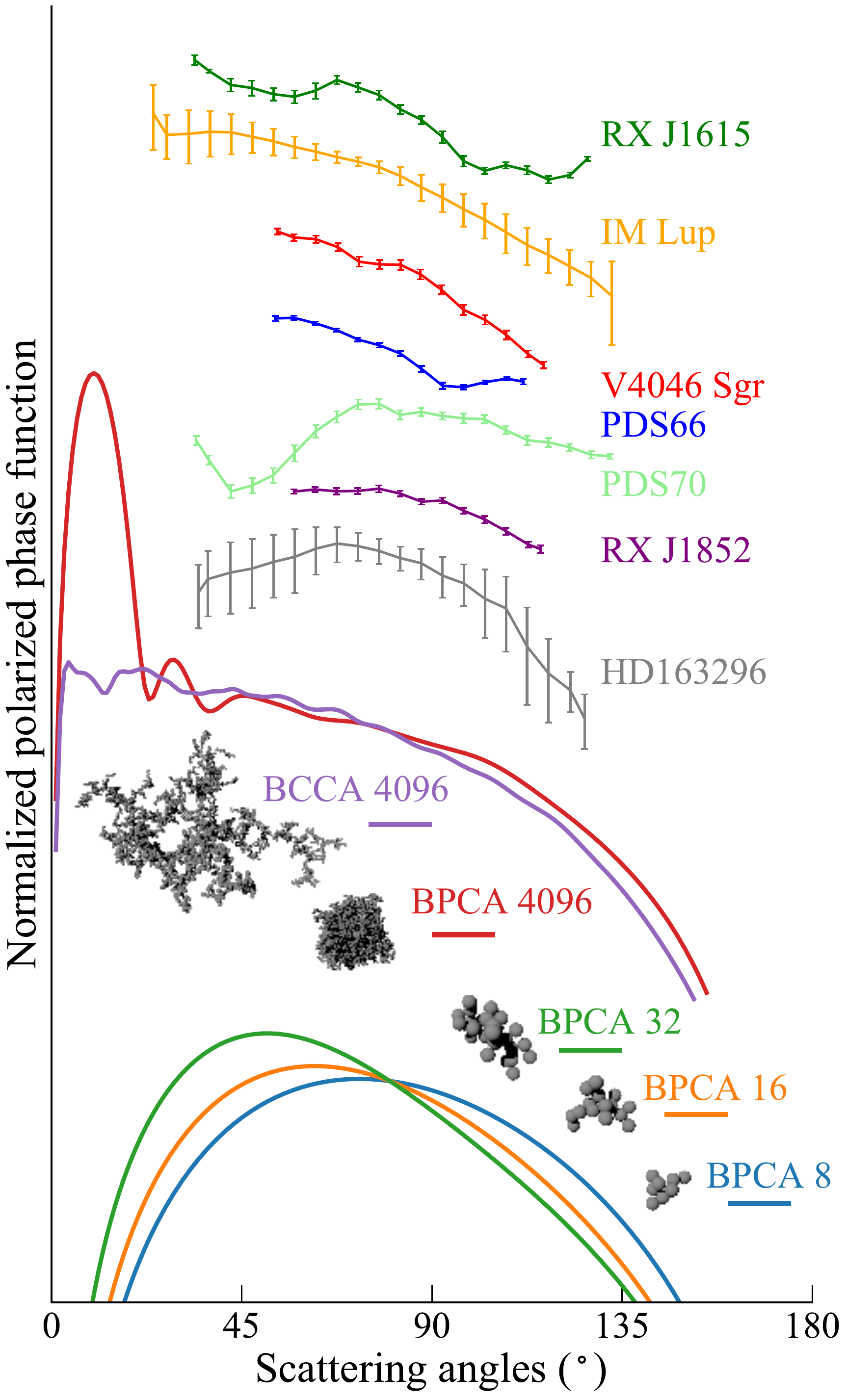}
\caption{\small Polarized phase functions extracted from $H$-band observations (lines with error bars; Ginski et al.\,in prep.), compared with the T-Matrix light scattering simulations (full lines; Tazaki et al.\,in prep). The upper set of theoretical curves correspond to dust aggregates consisting of 4096 monomers of radius of 0.1 $\mu$m with resultant aggregate radii of approximately 10 µm (BCCA; an open structure aggregate) and 3.1 µm (BPCA; a compact structure aggregate). The lower set of curves is for smaller dust aggregates consisting of 8, 16, 32 monomers with resultant aggregate radii of 3.3, 4.5, 6.0 µm, respectively.}
  \label{fig:phase_functions}
 \end{figure}

Most parts of the disks are optically thick at optical and near-IR light, and therefore, scattered light observations trace dust in the upper layers of disks, often several pressure scale heights above the \rev{midplane \citep{Chiang1997,Lagage2006}}.  \rev{The dust in these regions} should be well-coupled to the gas, with slow settling processes and efficient vertical mixing processes. \rev{This means that the dust should be either small individual grains \citep{Dong2012}, or porous, possibly fractal aggregates.} \rev{Models of dust aggregation generally show that aggregation is fast \citep{Dullemond2005,Okuzumi2012}, and that the velocities needed to destroy aggregates in collisions are only reached when these aggregates become large enough to start decoupling dynamically form the gas \citep{Birnstiel2010}, so there are good reasons to expect the dust in the disk surface to be present on the form of aggregates.  However, fragmentation and infall of molecular cloud material can replenish small individual particles to some degree, so observational tracers are needed determine the properties of the dust.  Here, we focus much of our discussion of dust properties of optical properties of aggregates.}

Information about the properties of the dust is encoded in the intensity and degree of polarization of the reflected light, in the angular dependence of these quantities (phase functions), and in the wavelength dependence \citep[e.g.][]{Takami2014}. Grains with sizes much smaller than the wavelength of the observations scatter isotropically, with a bell-shaped polarization fraction peaking around a scattering angle of 90\degree~(see Figure\,\ref{fig:scatgeometry}). The degree of polarization at the peak can get quite high, typically 20 to 70\% \citep{krivova2000,Silber2000,Graham2007,Perrin2009,Murakawa2008,Tanii2012,Poteet2018,Monnier2019,Hunziker2021,Tschudi2021} and should be considered a lower limit since multiple scattering or limited resolution would tend to reduce the measured degree of polarization. Larger grains concentrate more of the scattered light into a small angle around forward-scattering direction, with smaller degrees of polarization, still bell-shaped around 90\degree. Aggregates can behave like large grains when it comes to the amount of forward scattering \citep{Kozasa1993}. 

As discussed by \citet{tazaki2019}, aggregate sizes can be derived from the ratio of the forward and backward scattering intensity.  If the aggregates are larger than the observing wavelength, then the polarization fraction is a measure of the porousness of the aggregates, with higher polarization degrees indicating higher porosity. The color of total and polarized intensity images indicate dust properties as well, with porous aggregates leading to gray or slightly blue colors, while more compact aggregates cause reddish colors in total intensity \citep{Mulders2013a}.

\begin{figure*}[ht]
\centering
  \includegraphics[width=0.9\textwidth]{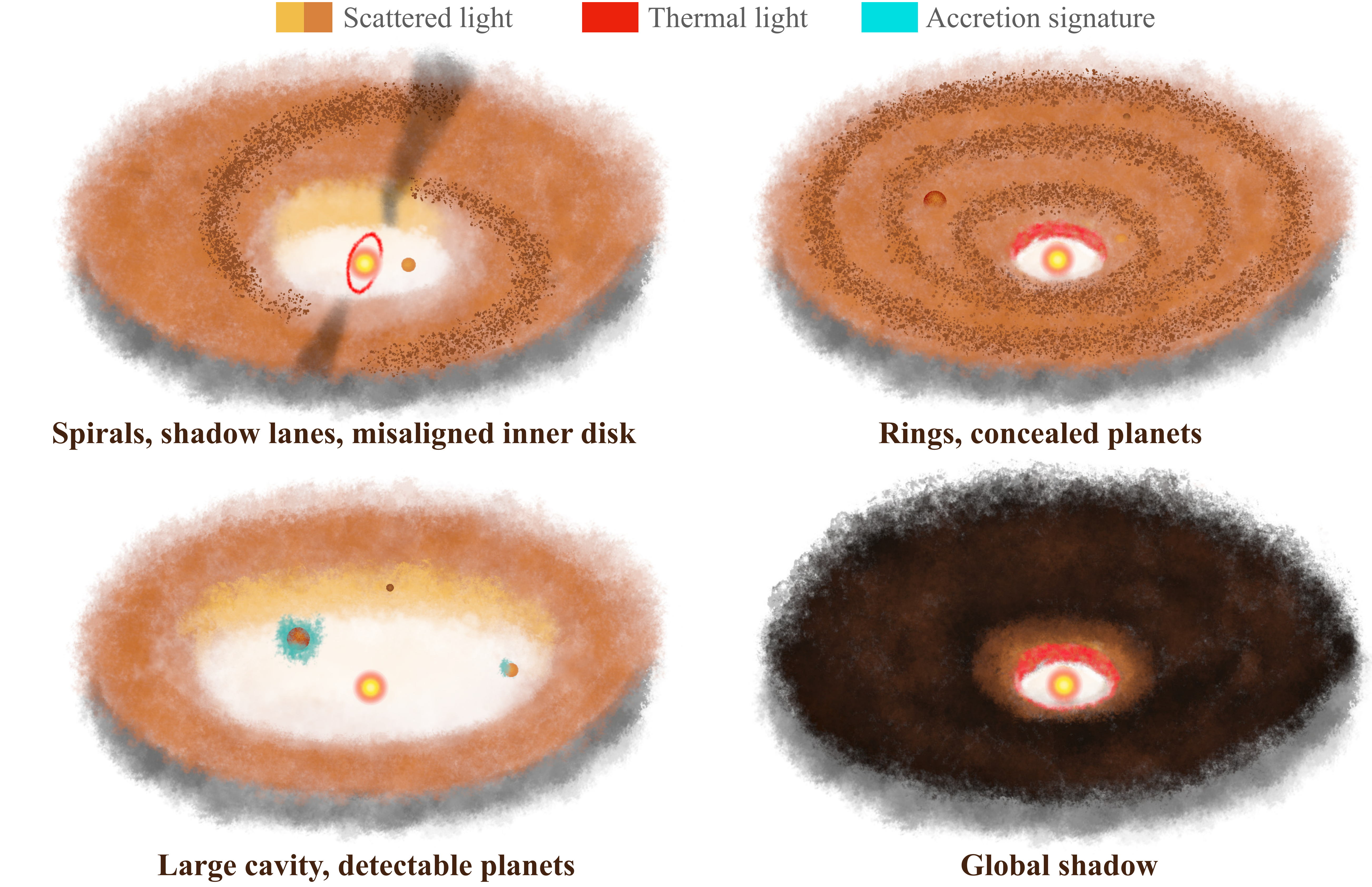}
  \caption{\small Visual synopsis of the topics treated in Sect.\,\ref{sec:main_substructure} and \ref{sec:main_protoplanet}. Disk substructures such as cavities, rings, and spirals are discussed in Sect.\,\ref{sec:substructures}, shadows and misaligned inner disk in Sect.\,\ref{sec:shadows}, while protoplanets and accretion signatures in Sects.\,\ref{sec:PDS70} and \ref{sec:protoplanets}.}
  \label{fig:sketch}
 \end{figure*}

\rev{For \textit{intensity phase curves}, many sources do indeed show significant forward scattering enhancements \citep[e.g.][]{Quanz2011,Hashimoto2012,Kusakabe2012,Mulders2013a,Ginski2016}.  When a full range from small to large scattering angles can be probed (see also see Figure~\ref{fig:scatgeometry}), in both protoplanetary and in debris disks, the phase function is usually characterized by a very strong forward scattering peak, a somewhat flat region around 90 degrees, and a rise toward backward scattering \citep[e.g.][]{Hughes2018,Milli2017,Stolker2016},  consistent with lab measurements of large, porous aggregates \citep{Munoz2017}.
\textit{Polarized phase functions} have now been measured for a number of disks, over significant ranges in scattering angle (Figure~\ref{fig:phase_functions}). Such polarized phase functions depend sensitively on aggregate radii. Smaller aggregates show a symmetric polarized phase function with respect to a scattering angle of 90$^\circ$. In contrast, larger aggregates tend to develop forward scattering component. Aggregate structure also affects how strongly the forward scattering peak is expressed.
Looking at the observations in Figure~\ref{fig:phase_functions}, there seem to be two classes of polarized phase functions. One is a smoothly increasing phase function for smaller scattering angles, such as IM\,Lup, V4046\,Sgr, RX\,J1852. This class of phase function shows a very strong resemblance to the polarized phase functions of large ($\gtrsim\mu$m) highly porous aggregates. The other class is a phase function with a strong drop-off at small scattering angles, such as PDS\,70, HD\,163296.  Here, the increase in scattering intensity is not able to compensate the decrease in polarization fraction, leading to a drop in the polarized phase functions toward these forward-scattering angles.  Such a functional shape can be well reproduces with very small aggretates ($\lesssim\mu$m), or perhaps larger but very compact aggregates/particles.}

\rev{The dust material can imprint its solid-state features in the scattered light as well. This should in principle give the most direct access to the grain composition, but this approach is limited by the small number of usable solid state features in the optical and near IR range. So far, scattered light displaying the water ice feature at 3$\mu$m has been detected in HD\,142527 \citep{Honda2009} and HD\,100546 \citep{Honda2016}.
A detailed study of the ice feature observed in HD\,142527 shows that the dust has an ice/core mass ratio between 0.06 and 0.2 \citep{tazaki2021}, where the core was assumed to be made of silicate but may also contain other components that do not show an absorption feature at that wavelength.}

\section{Substructures} \label{sec:main_substructure}
Since PPVI, near-IR and ALMA high angular resolution images have shown that disk substructures are very frequent, appearing as dust depleted cavities, rings, gaps, asymmetries and spiral arms. Figure~\ref{fig:sketch} is a sketch of such substructures and of the topics discussed in this section. 

\subsection{Cavities, rings, spirals} \label{sec:substructures}
\subsubsection{Cavities} 
The existence of large ($>10$ au) regions devoid of, or with significantly dimmed amount of dust grains was first predicted from the lack of near- and mid-IR excess from the SED \citep{Strom1989}. These objects were initially defined as transition disks \rev{(referred to as "transitional disks" in the PPVI review of \citet{PPVIEspaillat})} \rev{since it was believed that they were rapidly converting}  inside-out from a dust-rich to a dust-poor composition. The first resolved images \rev{were obtained for the extraordinarily large cavities of the multiple systems GG Tau \citep[with the IRAM interferometer]{Dutrey1994} and HD142527 \citep[with Subaru/CIAO]{Fukagawa2006}. Cavities are now routinely resolved by PDI observations \citep[e.g.,][]{Mayama2012, Hashimoto2012}}. However, in many cases the near-IR cavity is either smaller than when measured at millimeter wavelengths \citep[e.g.,][]{Garufi2013, Villenave2019, Mauco2020} or undetectable at radii down to coronagraph \citep[typically at 10--15 au; e.g.,][]{Benisty2015, Ginski2016, MuroArena2020}. 
This spatial segregation of dust size is currently best explained by the existence of dust trapping of mm-sized pebbles at a local pressure maximum induced by a planetary companion \citep{Pinilla2012}. In this scenario, gas still flows through the cavity and carries well-coupled small grains with it. Therefore, one naturally expects a radial segregation between small and large grains \citep{deJuanOvelar2013} that may be used to predict the position and the properties (mass) of a putative planetary companion responsible for the cavity (for a given disk viscosity). \\

\begin{figure*}[!t]
\centering
  \includegraphics[width=1.0\textwidth]{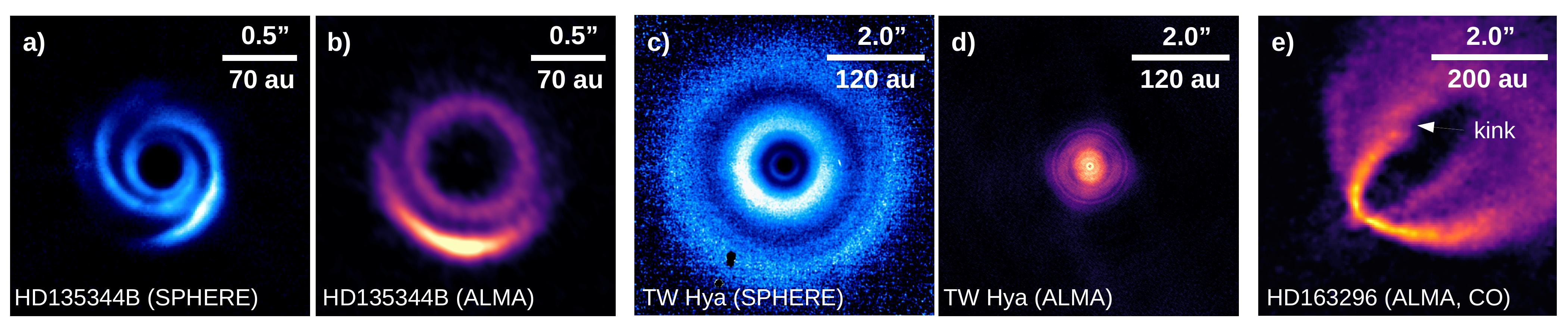}
  \caption{\small Comparison of IR scattered light images with ALMA observations. Panels (a)(b) for HD135344B, Panels (c)(d) for TW Hya. Panel (e) shows the detection of a non-Keplerian feature (so-called kink) in a channel map of the CO observations of HD163296.}
  \label{fig:alma}
 \end{figure*}

\subsubsection{Rings} 
\rev{Multiple} rings in PDI were first detected in the Herbig Ae star HD169142 with early NACO observations that revealed the presence of two rings \rev{\citep{Quanz2013b}}. Subsequent observations of the same disk with more advanced instrumentation indicated that the first ring is highly structured \citep{Pohl2017} with asymmetric clumps distributed along the ring that rotate at Keplerian rotation \citep{Gratton2019}. Multi wavelength scattered light observations in the J and H \rev{bands} can show a differential color for the two rings indicating that the stellar light hitting the outer disk has been reddened while traversing the first ring \citep{Monnier2017}. Subsequent observations of T Tauri and Herbig Ae stars indicated that multiple rings are quite common in T Tauri disks \citep[e.g.,][]{Rapson2015,Thalmann2016, Monnier2017,MuroArena2018, Avenhaus2018,Bertrang2018}.  Overall, rings are found at all radii accessible to direct imaging, that is from the inner working angle \citep{vanBoekel2017}, to the outermost radius probed at the given sensitivity \citep{deBoer2016}. Using analytical prescriptions for the gap morphology \citep{Kanagawa2015} or hydrodynamical simulations followed with radiative transfer \citep{Dong2017}, the properties of the \rev{annular} gaps (width, depth) are related to the presence of massive planets for a given disk viscosity which can then be compared to current detection limits \citep[see Sect.\,\ref{sec:protoplanets}]{Asensio-Torres2021}.
Other explanations are provided, for the presence of rings seen in the surface layers, for example ice lines \citep{Okuzumi2012} of different volatile species. They can perturb dust dynamics locally which has a direct observational consequence \citep{Pinilla2017}. \cite{Pohl2017} compared the location of the snow line for various volatile species and found that for HD169142, the outermost gap could be explained by dust accumulation close to the CO ice line. This study has unfortunately not been done on a larger sample of rings and a morphological and statistical study of the rings properties in a large survey of T Tauri stars, in terms of spatial distribution, width, and contrast still remains to be done to infer whether they have a similar origin. 

Many of the disks that show \rev{multiple} rings in the sub-millimeter also exhibit rings in IR scattered light (see Chapter \textit{Bae et al.}). However, there is no general correspondence between the location of the rings in the sub-millimeter and in the IR, nor in the numbers of rings detected. One of the clearest example is TW Hya \citep{vanBoekel2017, Andrews2016}, as showed in Figure\,\ref{fig:alma}, panels c and d, not only the radial extent of the small grains distribution (seen in IR) differs strongly from the one of the large grains (seen with ALMA) but the \rev{annular} gap/rings location do not match. This, in addition to providing a textbook example of radial drift in disks, shows the difference in dust/gas coupling and that the disk can be highly structured beyond what is evident from sub-millimeter images. \cite{MuroArena2018} showed that in the case of HD163296, only one of the two rings imaged by ALMA \citep{Isella2018} is detected in VLT/SPHERE observations. As the second, further, ring is not detected in scattered light, a natural explanation is that it lies in the shadow of the first ring, and must be significantly settled. These studies show that the analysis of a single dataset, obtained within a single wavelength regime, provides limited information and that it is crucial to perform a multi-wavelength analysis of these disks to constrain the mechanisms at play. Such a multi-wavelength approach can for example address the efficiency of dust evolution processes (radial drift, dust settling), as showed by the modeling of highly inclined disks at these two wavelengths \citep{Villenave2019,Villenave2020}. \\


\begin{figure*}[!t]
\centering
  \includegraphics[width=1.0\textwidth]{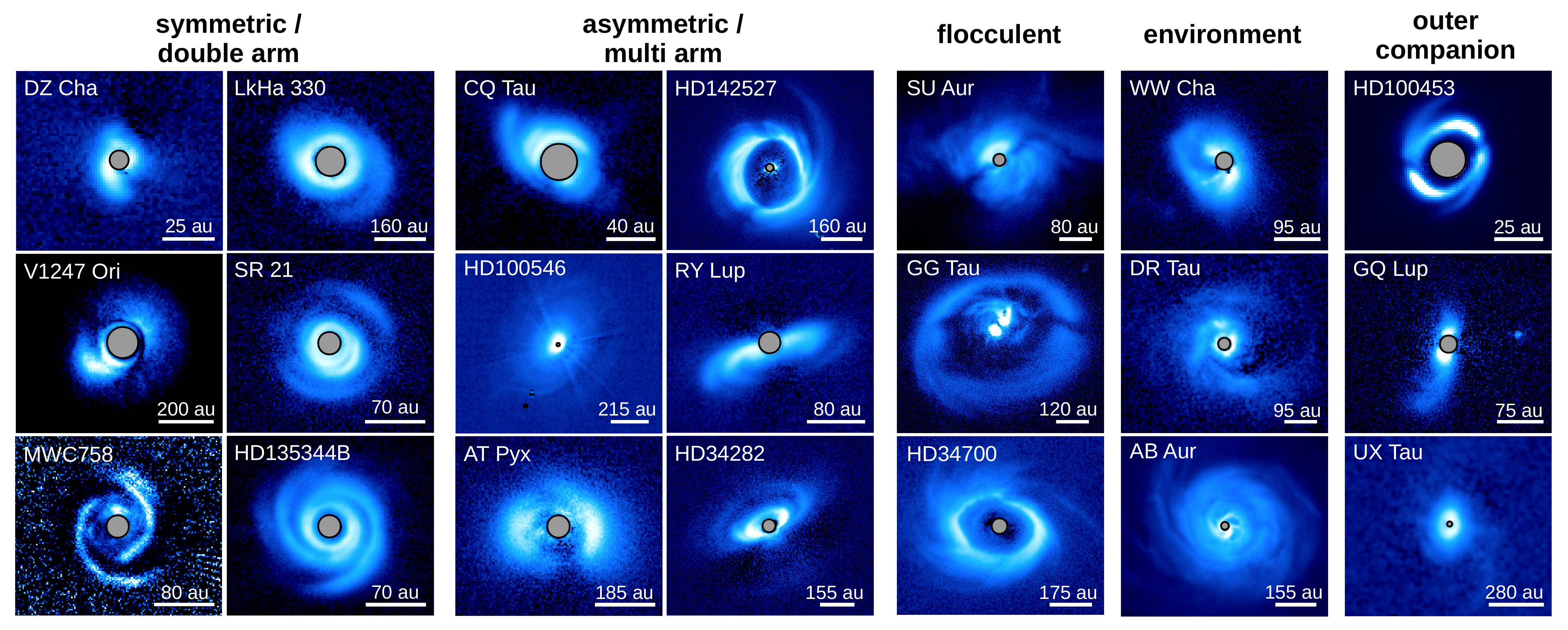}
  \caption{\small Spiral features imaged in scattered light at the time of this review. \rev{The majority of the data was taken with VLT/SPHERE with the exception of HD\,34700, which was imaged with Gemini/GPI and LkHa\,330 and V1247\,Ori which were imaged with Subaru/HiCIAO. All relevant references for individual systems are listed in appendix~\ref{App: ref: spiral gallery}. All data were taken in the IR J and H-bands with the exception of LkHa\,330 and MWC\,758 which were imaged in the K and Y-band respectively.} Color bars were individually adjusted to highlight the disk morphology. Categories are based on the spirals appearance alone, as the driving mechanism behind the spirals is still unclear (\rev{we do however list disks with a known outer companion in a separate category). We note that for several disks in principle multiple categories apply, e.g. AB\,Aur clearly still interacts with the large scale cloud environment, but does also show multiple spiral arms or HD\,100453 has an outer companion but is also a prime example for a symmetric double-armed spiral disk. Without the ability to disentangle driving mechanisms such a categorization by appearance will always contain a certain amount of subjectivity.}}
  \label{fig:spiral-gallery}
 \end{figure*}

\subsubsection{Spirals}  Figure\,\ref{fig:spiral-gallery} provides images of disks presenting spiral features, that can  be separated into various categories based on their morphology and on their possible origin. 
Some disks show two nearly symmetric arms \citep[e.g.,][]{grady2013,Wagner2015,Benisty2015,Stolker2016a}.
\cite{Juhasz2015} showed that these spirals are likely tracing local perturbation of the disk scale height, rather than the density perturbations, and that in general opening angles are too large to be accounted by a companion located inwards of these spirals. This issue can be solved by considering a massive companion located beyond the spiral arms in the outer disk outside of the disk \citep[][]{Dong2015}. However, these massive planets have not been detected. Recently,  \citet[][]{Calcino2020} found that a 10\,M$_{\rm{Jup}}$ planet orbiting within the cavity on an eccentric orbit could account for the properties of the spiral arms. 

Other disks show multiple asymmetric arms \rev{that may sometimes appear flocculent} \citep[e.g,][]{Monnier2017}. While it is unclear whether this can be the origin in all cases, some of these targets are known circumbinary disks for which spiral arms are triggered by the interaction of the disk with the stellar system within \citep[e.g.,][]{Price2018, Monnier2019}. In other cases, e.g., HD34282 or CQ Tau \citep[e.g,][]{Uyama2020,deBoer2021}, the overall disk surface geometry might lead to apparent complex structures. Impressive, complex, and multiple spiral arms are observed in systems that show clear interaction with the surrounding material, possible remnant of envelope \citep[as in e.g., \rev{AB Aur and DR Tau,}][]{Boccaletti2020, Mesa2022}, and for which the synergy with ALMA observations is once more essential. In SU\,Aur, for example, the geometry of the infalling material could be determined thanks to the combined analysis of IR scattered light and ALMA gas tracers \citep{Ginski2021}. Another category includes large scale spiral arms that are due to the dynamical interaction of the disk with a \rev{known} external companion as in the SR24 and UX Tau systems \citep[e.g.,][]{Mayama2020, Menard2020}, possibly resulting of a recent fly-by event. 
\rev{We note that the classification proposed in Figure \ref{fig:spiral-gallery} is not absolute, and that several systems may fit in multiple categories. We attempted to classify based on the most prominent features, e.g. AB\,Aur shows large scale interaction with the surrounding cloud and is cited as candidate for possible late material infall by \cite{Dullemond2019}, however it also clearly shows multiple spiral features and could thus also be categorized as multi-armed spiral disk.}

Recently, \citet{Ren2020} discussed that multi-epoch imaging can be used to discern small proper motions in spiral arms.  Different physical causes for spirals and blobs make clear predictions for the proper motions, since some structures might have a 'pattern' speed that is lower than the orbital timescale.  For instance, \citet{Ren2020} detect a small rotation ($\sim0.22\arcdeg/\mathrm{yr}$) of the MWC~758 spiral pattern,  slower than the local Keplerian motion expected if spiral structure derived from gravitational instability; instead, a perturber located in the outer disk is hypothesized.  Similarly, slight spiral motion reported by \citet{Xie2021a} for HD135344B point towards external perturbers. 
We are only just beginning to accumulate results from proper motion studies and we can expect a rapid expansion as the time base between epochs continues to increase.  

Similarly to the rings, only a handful of disks show a clear correspondence between spirals observed in the IR and in the millimeter continuum, indicating the complexity of disk evolution processes \citep{Dong2018,Brown2021}.  For example, AB Aur shows large scale multiple spiral arms in scattered light \rev{\citep{Fukagawa2004, Hashimoto2011, Boccaletti2020}} and, in contrast, a ring in the sub-millimeter continuum \citep{Tang2017}. Another case is presented in Figure\,\ref{fig:alma}. HD135344B shows two spiral arms in the IR \rev{\citep[e.g.,][]{Muto2012}}, while in the sub-millimeter regime, presents a ring with an additional azimuthal asymmetry at larger radii \rev{\citep{vanderMarel2016b, Cazzoletti2018b}}. Such a correspondence is often observed \rev{\citep{Garufi2018, vanderMarel2021}}. 
While joint analysis of high resolution images at both wavelengths has only been done in a handful of cases \citep{Dong2018,Baruteau2019},  it has the potential to constrain the possible location of the planet/companion, the 3D structure of the disk and in particular, its vertical thermal structure. \cite{Rosotti2020} compared the morphology of spirals of HD100453 observed with SPHERE and ALMA, and found that the opening angles of the spirals in the surface layers is much larger than in the midplane tracer. As the opening angle is directly related to the sound speed, and hence to the disk  temperature, \cite{Rosotti2020} interpreted this as a clear evidence for a vertical thermal stratification. 




\begin{figure*}[t]
    \centering
    \includegraphics[width=1.0\textwidth]{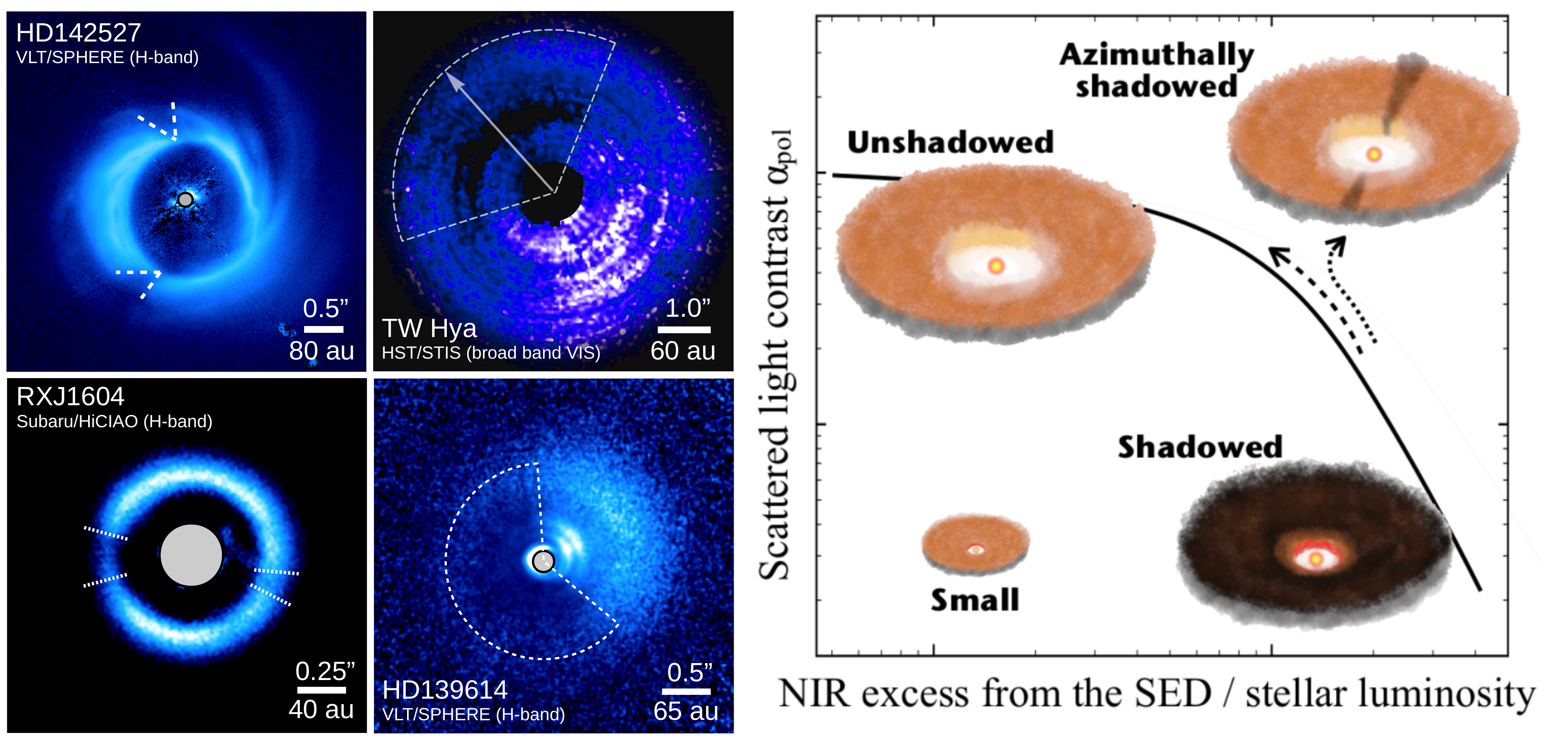}
    \caption{\small \rev{Shadows in disks. \textbf{Left}:
    Four exemplary cases of azimuthal shadows detected in disks with different facilities. The disks around HD\,142527 (\citealt{Hunziker2021}) and RX J1604 (\citealt{Mayama2012}) show narrow shadows, indicative of a strong misalignment of inner and outer disk. TW\,Hya (\citealt{Debes2017}, image credit: ESA) and HD\,139614 (\citealt{MuroArena2020}) show broad shadows, indicative of a small misalignment or warp in the disk. \textbf{Right}: Sketch illustrating the interplay between inner and outer disk, observed as an anti-correlation between the disk brightness in scattered light and the near-IR excess \citep[black solid line, see][]{Garufi2022}.} 
    }
    \label{fig:shadows}
\end{figure*}

\subsection{\rev{Azimuthal} shadows \& misaligned inner disks} \label{sec:shadows}
Shadows are another recurrent feature of disk images in scattered light. \rev{While in principle any region of low surface brightness seen in polarimetric maps can be due to shadows, the community widely names shadows only those} dips in the disk brightness confined to specific azimuthal angles. \rev{Here we refer to these structures as azimuthal shadows.} The prototypical example of such features is the two intensity drops seen to north and south in the disk of HD142527 \citep[see Figure \ref{fig:shadows}]{Canovas2013, Avenhaus2014a}.  Through radiative transfer modeling, \citet{Marino2015} interpreted these regions as shadows cast by a significantly misaligned portion of the inner disk. Such a disk component was possibly imaged by \citet{Avenhaus2017} as close as 4 au from the star. In general, \rev{azimuthal} shadows with a diverse morphology were revealed in several disks, as we review in this section and \rev{show in Figure~\ref{fig:shadows}.} These features offer the possibility to indirectly map the inner disk geometry and its temporal evolution \rev{since} marked changes in the scattered light emission on months to years times scales can be caused by shadowing from the inner disk with its shorter dynamical timescales.  Further, distinctive shadow patterns tell us about the inner disk geometry on sub-au physical scales typically only accessible by long-baseline IR interferometry.

\begin{figure*}[t]
    \centering
    \includegraphics[width=0.90\textwidth]{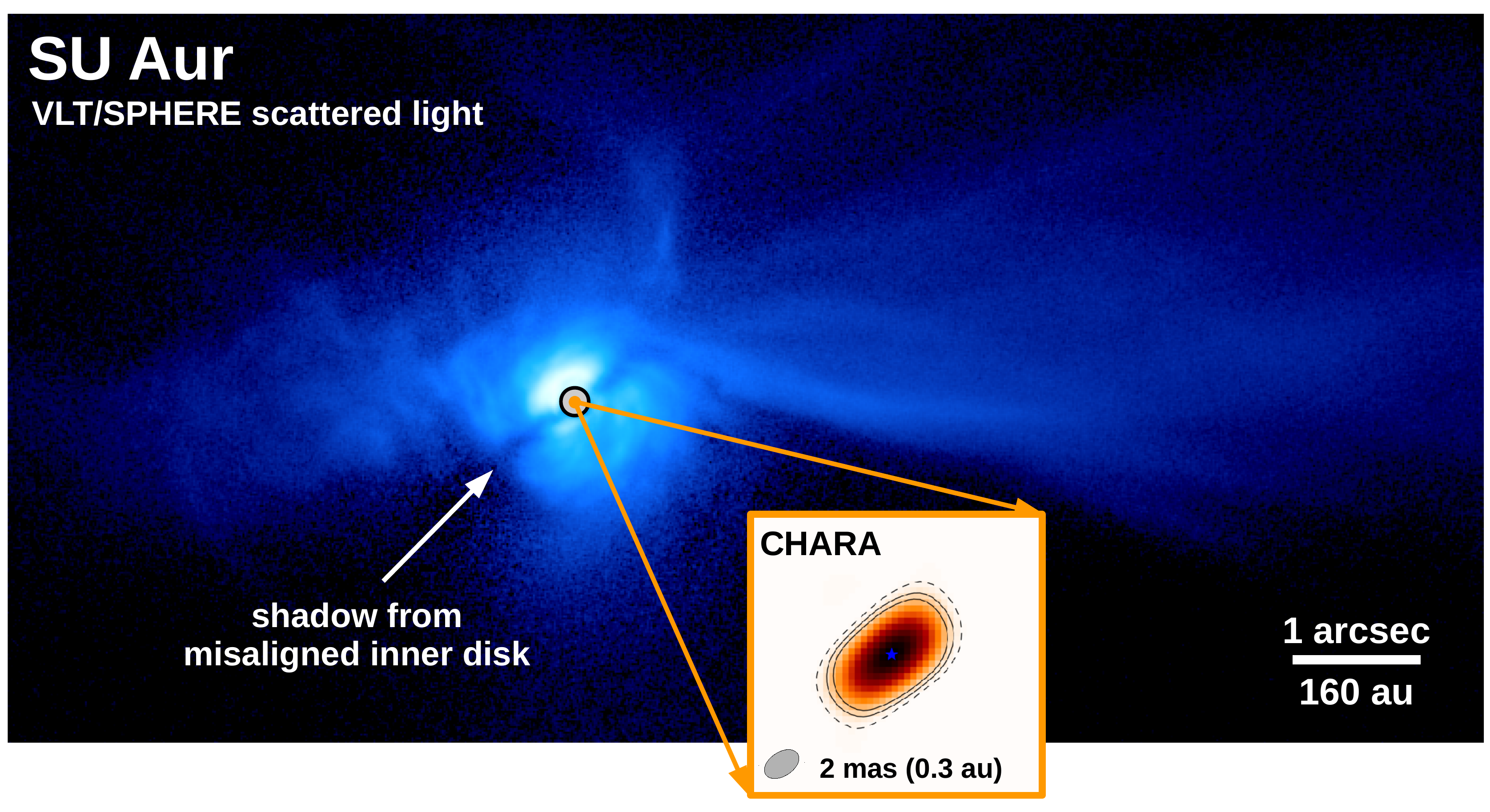}
    \caption{\small Misaligned disk system in  SU~Aurigae. A complex scattered-light disk is seen on large scales \citep[VLT/SPHERE;][]{Ginski2021} with a dark lane caused by shadowing from a misaligned inner disk that has been directly imaged with the CHARA Array \citep{Labdon2019}. The inset box is zoomed by a factor of $\sim$1000$\times$.
    }
    \label{fig:suaur}
\end{figure*}

Two main classes of \rev{azimuthal} shadows are recurrently observed, that are narrow shadows of a few degrees in azimuthal extent that are often seen as two symmetric narrow lanes, and broad shadows where up to half a disk is faint (see Figure\,\ref{fig:gallery} and Figure\,\ref{fig:shadows}). A particularly clear example of narrow shadow can be found in SU~Aur. In Figure~\ref{fig:suaur}, the outer disk and envelope are illuminated in scattered light \citep{Ginski2021}.  Near the central star, a dark lane can be seen suggesting a shadow cast by a physically-thin disk that is very misaligned with the outer disk.  Indeed, in this rare case an near-IR interferometry image on 1000$\times$ smaller scales was obtained using the CHARA Array \citep{Labdon2019} confirming this hypothesis.  In the case of SU Aur, it is possible that such a misalignment is caused by late-time interactions with infalling material \citep[e.g.,][]{Dullemond2019}. 
Other notable examples of narrow shadows include GG Tau A \rev{\citep{Krist2002, Itoh2002, Itoh2014, keppler2020}} -- where the shadows are due to the material in the surrounding of the binary system, RX J1604.3-2130 \citep{Pinilla2015, Pinilla2018b} and HD135344B \citep{Garufi2013, Stolker2017} -- where the dips are highly variable in both morphology and position, HD100453 \citep{Wagner2015, Benisty2017} -- where shadows appear morphologically connected with the spiral arms, and DoAr44 \citep{Avenhaus2018} -- where the shadows have a radio counterpart \citep{Casassus2018}. Recently, \citet{Bohn2021} analyzed a sample of 20 transition disks around T Tauri and Herbig AeBe stars, observed with VLTI/GRAVITY and ALMA to search for evidence for misalignments between the inner and outer disk regions. They found that six disks exhibit significant misalignments, among which three disks (HD100453, HD142527 and CQ\,Tau) show shadows in scattered light images, with locations that match well the predictions from the inner/outer disk geometry derived with VLTI and ALMA respectively. For three other disks that are found significantly misaligned (V1247~Ori, V1366~Ori, and RY~Lup), there is no evident sign of shadows in the scattered light images. On the other hand, the shadows observed in DoAr44, HD135344~B, and HD139614 while consistent with misalignments show a complex morphology that can \rev{not} be well described with a simple misaligned inner disk. 

In general, when there is smaller misalignment, an inner disk will cast a broader, less distinctive, shadow. HST/STIS \citep{Debes2017} found a moving shadow pattern around TW~Hya, with a timescale suggesting an precessing inner disk (possibly driven by a close-in exoplanet).  \citet{Benisty2018} found a dramatic contrast between the disk surface brightness on the east compared to west side of HD~1430006 that can be modelled by an inner disk misaligned by 30$\arcdeg$. As explained in Section\,\ref{sec:theory}, an inner massive exoplanet on a slightly inclined orbit can sufficiently misalign the inner disk and cast similar wide-angle shadows \citep{Nealon2019, ballabio2021}.   At first glance, HD~139614 appears to be another case of a nearly-aligned inner disk casting a wide-angle shadow \citep{MuroArena2020,Laws2020}, but the shadow is {\em too wide} and \citet{MuroArena2020} argue for the existence of two distinct warped or misaligned regions. Note that even when the inner disk is mostly aligned, there can be more subtle shadowing effects such as the reddening of starlight (not full shadowing) by inner dust rings that partially occlude the outer disk \citep[e.g., HD169142;][]{Monnier2017}.

Not all shadow patterns are caused by misaligned disks.  Even within the misaligned-disk paradigm, many workers \citep[e.g.,][]{Stolker2017,Pinilla2018b} noted additional shadow variability requiring scale-height (H/r) variation in the inner disk atmosphere that could be due to irregular clumps or dust lifted at the base of magnetospheric accretion columns. HD~163296 shows \rev{globally} variable scattered light emission \citep{Rich2019,Rich2020} but we know the inner and outer disks are well aligned \citep{Monnier2017,Setterholm2018}.  \citet{Rich2019} suggests obscuration by dust clouds elevated in a disk wind might be at play while recent VLTI/MATISSE and VLTI/GRAVITY interferometric observations of the HD~163296 reveals an inner disk asymmetry possibly caused by a vortex \citep{Varga2021,Sanchez2021}.  There is a likely a close connection between these shadowing mechanisms to the well-studied photometric "dipper" phenomena \citep[e.g.,][]{Cody2014,Bodman2017} where young stars show diverse light curves due to variable obscuration by intermediate and high inclination close-in disks.  The causes for such variations in scale height could include magnetic field interactions \citep{Bouvier1999}.   

Supporting this general picture, \citet{Garufi2018} discussed an empirical correlation between disks with narrow shadows and anomalously high \rev{thermal} near-IR excess from the SED, possibly indicating a dynamically perturbed inner disk. Following up on this finding, \citet{Garufi2022} found that, excluding these radially shadowed disks, large disks present an anti-correlation between the disk brightness in scattered light and the near-IR excess. Therefore, a final picture is emerging for the inner/outer disk interplay as is drawn in the sketch of \rev{Figure\,\ref{fig:shadows} (right panel)}, \rev{with the inner disk geometry determining the illumination pattern in the outer disk. On the one hand, uniformly shadowed disks (see Sect.\,\ref{sec:brightness}) may in principle evolve as unshadowed disks as the material in the inner disk is depleted or as the inner disk vertical structure changes. On the other hand, radially shadowed disk may represent a particular stellar-companion-disk configuration that leads to an inner disk geometry that yields both an exceptionally high \rev{thermal} near-IR excess and azimuthally confined shadows.}




Time variability of the shadow patterns themselves require either precession of the inner disk with respect to the outer disk, or orbital evolution of inner disk warps or height structure in the inner disk wall. While inner disk precession is likely slow ($\gg$ orbital time), Keplerian rotation of warps or inner disk structures will cast long shadows that can change on months timescale.    Note that the changing shadowing also changes the temperatures in the upper layers and \citet{Espaillat2011} reported ''see-saw'' changes in IR SEDs of T Tauri disks using Spitzer, observing increased near-IR flux (e.g., higher scale height of inner disk) associated with decreased mid-IR flux (more shadowing by inner disk). The variability of the shadows position and contrasts observed in RX J1604.3-2130 was found to be within timescales of less than a day \citep{Pinilla2015}. This implies that the dust casting the shadow is located very close to the star (0.06 au, at corotation radius), in an irregular misaligned inner disk, which dust content is accreted/replenished by accretion in very short timescales \citep{Sicilia2020}. \rev{In contrast, the shadow detected in TW Hya was measured to move at a constant angular velocity of 22.7$^\circ yr^{-1}$ in a counterclockwise direction, corresponding to a period of 15.9 yr assuming circular motion  \citep{Debes2017}. 
\rev{The sample of disks for which the temporal evolution of shadow features is being systematically tracked is still small, with currently only four cases in the literature. Two of them are the aforementioned cases of TW\,Hya as well as RX J1604.3-2130. In addition this was also done in the cases of HD\,135344B (\citealt{Stolker2017}) and PDS\,66 (\citealt{Schneider2014, Wolff2016}). The shortest monitoring time scales vary between days for RX J1604.3-2130, weeks for HD\,135344B, months for PDS\,66 and years for TW\,Hya. }}

\subsection{Demography of disk substructures} \label{sec:statistics}
At time of writing, nearly 130 stars hosting a planet-forming disk have polarimetric observations published. Out of these, the disk is firmly resolved around approximately 80 targets, with the remaining 50 equally distributed between non-detections and detection of scattered light from the environment. From the 80 detected disks, substructures are detected in 45 objects. For the remaining 35 disks, the absence of substructure can be explained by the small disk extent \citep[e.g., ET Cha,][]{Ginski2020},  the faint signal in scattered light \citep[e.g., CI Tau,][see Sect.\,\ref{sec:brightness}]{Garufi2022}, or the very inclined geometry of the disk \citep[e.g., T Cha,][]{Pohl2017}. Therefore, the current census indicates that disk substructures are ubiquitous when the disk is bright and extended enough. 

A demographical study of disk substructures in scattered light was carried out by \citet{Garufi2018}, who considered the 58 targets available at that time. All targets were classified in six main categories: disks hosting \rev{multiple} rings, one or two bright spirals on small scale, multiple gigantic spirals on large scale, a unique bright rim, as well as faint and small disks. 

\begin{figure}[ht]
    \centering
    \includegraphics[width=0.5\textwidth]{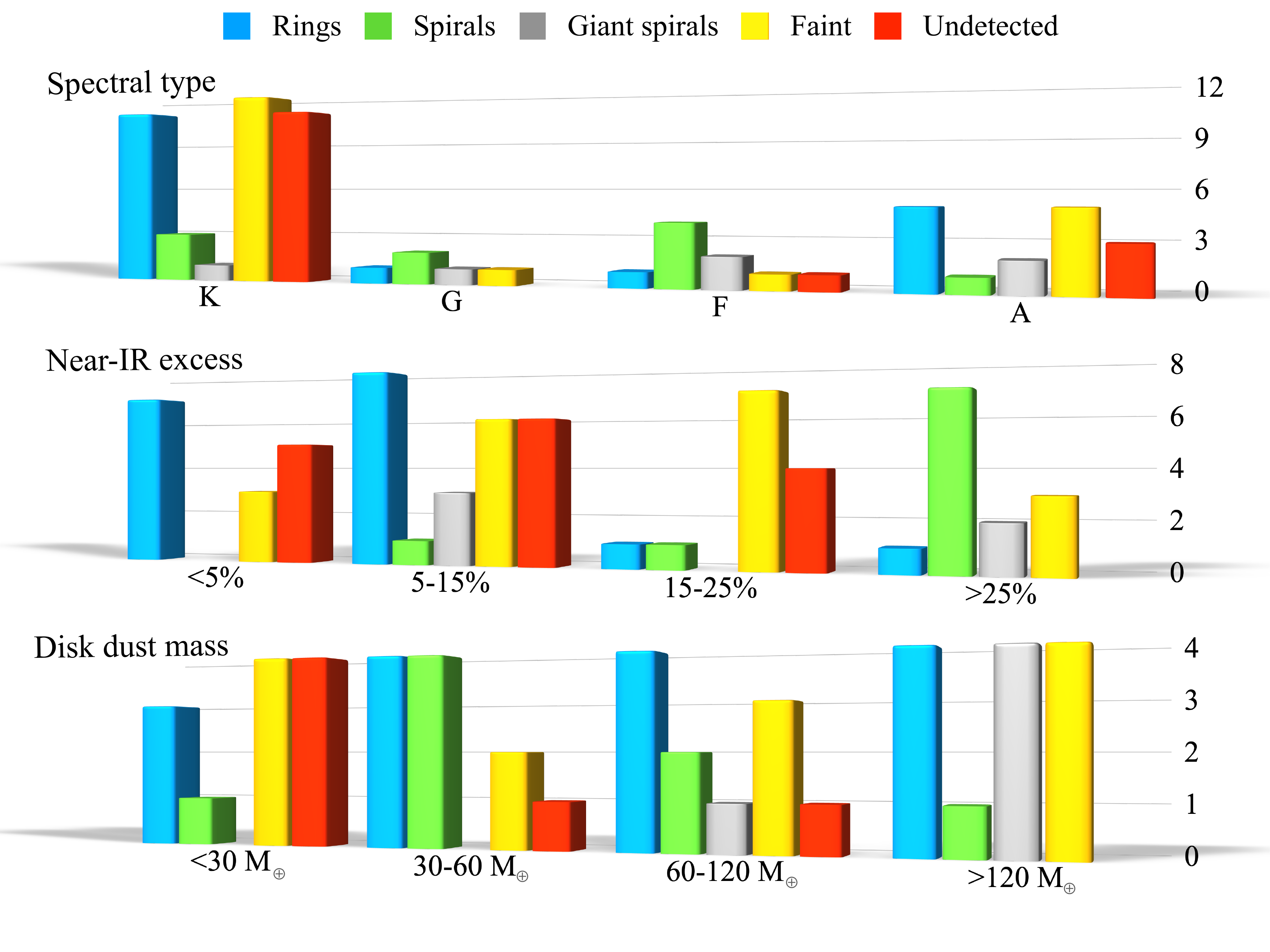}
    \caption{Distribution of targets with ring-, spiral-, giant spiral-, faint, and undetected disks with spectral type (top), near-IR excess (middle), and disk dust mass (bottom).}
    \label{fig:trends}
\end{figure}

At present, we can extend this type of study to a total sample of 76 objects excluding the aforementioned inclined disks and sources dominated by ambient emission. From our count, spiral arms are currently detected in 16 disks (6 giant disks with multiple arms at large scale and 10 smaller disks with 1-2 bright spirals at small scale), \rev{multiple} rings in 18 disks, and bright rim at the outer cavity edge in only 3 disks. In Figure\,\ref{fig:trends}, the distribution of the various classes of disks with three meaningful stellar and disk properties reveals trends in line with the results of \citet{Garufi2018}. 

\textit{Spiral disks.} Spirals are predominantly found around early-type stars. The recent discovery of spirals in the disk of WW Cha \citep{Garufi2020a} and GQ Lup \citep{vanHolstein2021} slightly changed this view. However, their formation is -- unlike those of isolated early-type stars -- probably due to the interaction with the accreting material and the stellar companion, respectively. The paucity of spiral-disks around late-type stars can in principle be explained by their low luminosity, and therefore by the low disk temperature because the detectability of spirals increases with the pitch angle which in turn is correlated with the aspect ratio and therefore temperature \citep{vanderMarel2021}, 
or alternatively by their low mass and young age. Disks with 1--2 spirals on small scale (tens of au) are not particularly massive whereas those with multiple spirals at large radii (hundreds of au) are. This suggests a possible different origin for these two classes of spirals. The origin of spirals may be intimately related to the structure of the inner disk since the near-IR excess from the SED of these spiral-hosting targets is significantly higher than that of ring-hosting targets.  

\textit{Ring disks.} Targets with ring disks are uniformly distributed across stellar properties and disk mass. However, the vast majority of disks with rings is found in sources with low near-IR excess. In this regard, disks with rings and spirals therefore show a dichotomic behavior.

\textit{Faint and undetected disks.} A very large number of targets with faint or no detection in scattered light is found around T Tauri stars. However, this behavior is not dictated by their low luminosity \citep{Garufi2022}. Instead, it is likely due to the young age of these sources. In fact, low-mass stars that are older than 3--5 Myr are currently not observed (see Figure\,\ref{fig:disks_imaged}) while the disk around intermediate-mass stars older than that can only persist if disk substructures such as an inner cavity is present, and this determines their brightness in scattered light (see Sect.\,\ref{sec:brightness}). Undetected disks are mostly those with low mass (since small) while several faint disks are found among those with high mass (since self-shadowed). 

\textit{Shadows}. The distribution of shadows with the spectral type is uniform from K to F types. However, none have been found around stars earlier than A9. As commented in Sect.\,\ref{sec:shadows}, shadows are -- similarly to spirals -- primarily associated with a high near-IR excess from the SED. This is particularly true for narrow shadows, as five of seven targets with such features sit in the upper quartile shown in Figure\,\ref{fig:trends}. So far, this empirical correlation is poorly explained by modeling effort but it clearly reinforces the connection between the morphology of shadows and that of inner environment around the star. 


\begin{figure*}[t]
    \centering
    \includegraphics[width=1.0\textwidth]{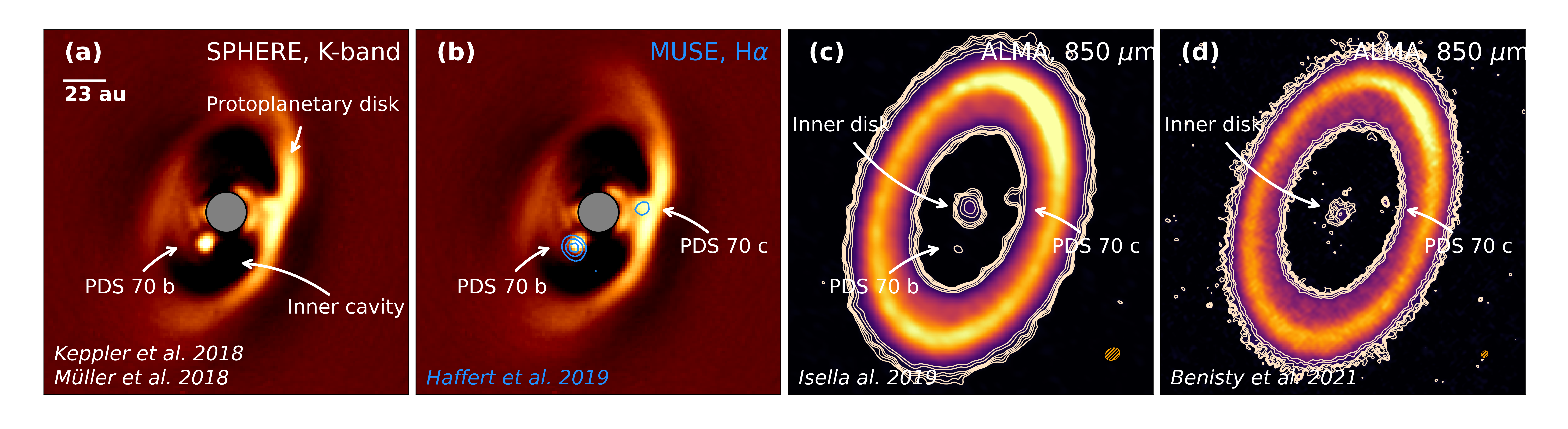}
    \caption{\small The PDS\,70 system as seen with SPHERE in the near-IR (left), overplotted with blue contours from H$\alpha$ observations (center left), and as observed in the sub-millimeter continuum (middle right, right). The difference between panels (c) and (d) is the angular resolution of the data, the beam is provided \rev{in orange color} in the bottom right corner. \rev{The SPHERE K-band data and the MUSE H$\alpha$ data both have an effective resolution of 70\,mas, i.e. a factor $\sim$3 worse than the ALMA data shown in panel (d). }}
    \label{fig:pds70}
\end{figure*}

\section{Direct imaging of planets} \label{sec:main_protoplanet}
High-angular resolution observations at near-IR and optical wavelengths are not only excellent tools to probe the properties of protoplanetary disks, but are also used to directly search for substructure-inciting planets within disks (see details in Chapter 22 by \textit{Currie et al.} on direct imaging and spectroscopy of extrasolar planets). 

Young, giant planets are still radiating remnant energy from the formation process, and direct near-IR thermal emission is one way to detect them (see Section~\ref{sec:protoplanets}). The relative contributions of direct photospheric emission and emission from circumplanetary material to the thermal emission from protoplanets is a subject of debate, and there is both observational \citep{Stolker2020,wang2021} and theoretical \citep{2015ApJ...799...16Z, 2019MNRAS.487.1248S, Szulagyi2021} support for substantial contribution from circumplanetary disks (CPDs). As the degree of CPD contamination of thermal emission from protoplanets is not well constrained, protoplanet mass estimates from evolutionary models carry a high degree of uncertainty. 

Additionally, \rev{some} transition disk host stars have accretion rates high enough to require the transport of gas from the outer disk \rev{\citep{Manara2014}}, through their cleared central cavities, and onto the star. Whether accretion onto the primary stars happens via high velocity low optical depth flows \citep[e.g.][]{2014ApJ...782...62R} or dense high-optical depth streamers \citep[e.g.][]{2013Natur.493..191C} remains a subject of debate, but either mechanism results in similar predictions of persistent accretion onto any protoplanets embedded inside transitional disk cavities. This has prompted several searches for accreting protoplanets in planet-forming disks \citep[see Section~\ref{sec:protoplanetsHalpha};][]{2020A&A...633A.119Z, 2019A&A...622A.156C}.  Accretion shocks caused by infalling gas on the planet itself or on the surface of the circumplanetary disk will give rise to \rev{both excess UV continuum emission from the stellar hotspot and line emission from infalling gas, making star-to-planet contrast ratios more favorable at a range of wavelengths where substantial accretion luminosity is present (see section \ref{sec:protoplanetsHalpha})}.

PDS 70b's detection in multiband thermal emission \citep[e.g.][]{Keppler2018} as well as in H$\alpha$ \citep[e.g.][]{Wagner2018,Haffert2019} and the lack of any potentially confounding nearby dust structures made it the first uncontroversial direct detection of an accreting protoplanet, as detailed in Section \ref{sec:PDS70}.  The successful operation of ALMA's longest baseline modes has enabled two complementary approaches to direct protoplanet detection. The first technique is to search for circumplanetary material, analogous to the disk hypothesized to exist around Jupiter in the protosolar nebula \citep{batygin13, Batygin2020}. This has been successfully applied in the case of PDS 70~c \citep{Isella2019,Benisty2021}, and the DSHARP survey \citep{Andrews2018} has also provided detection limits for CPDs in many observed mm \rev{gaps} \citep{Andrews2021}. The second method used for protoplanet study with ALMA is to search for kinematic signatures in mm gas emission, indicative of departure from Keplerian rotation due to an embedded planet (as detailed in section \ref{sec:kinks}).

\subsection{Protoplanets in the PDS70 system} \label{sec:PDS70}
As the most robustly confirmed case to date, the PDS\,70 system is an ideal laboratory to observationally study planetary system formation. At a distance of 112.4\,pc \citep{gaiaDR3} and with an age of $\sim$5 Myr \citep{Mueller2018}, PDS\,70 is a young T Tauri star surrounded by a protoplanetary disk. The system has long been suspected to host forming planets based on its large observed central cavity \citep[0$\farcs$43,][]{Long2018,Dong2012,Hashimoto2012,Hashimoto2015}. Multi-epoch near-IR observations using the VLT/SPHERE instrument \citep{Beuzit2019} as part of the SHINE exoplanet survey \citep{Chauvin2017} revealed a point source, PDS\,70\,b, at a projected separation of $\sim$200 mas and a position angle of $\sim$150\degr \ \citep[see Figure \ref{fig:pds70}, left;][]{Keppler2018,Mueller2018}. The planet was subsequently confirmed through the detection of accretion excess emission in the H$\alpha$ line and in UV continuum \citep{Wagner2018,Haffert2019,Hashimoto2020,Zhou2021}. H$\alpha$ imaging by \cite{Haffert2019} revealed a second accreting planet in the system, PDS\,70\,c, located at a projected separation of $\sim$230 mas and a position angle of $\sim$270\degr~(see Figure\ref{fig:pds70}, center panel).
While extraction of signal from PDS\,70\,b benefits from its location well within the gap of the disk, excluding any possible confusion with disk features,  PDS\,70\,c's emission is more challenging to disentangle from the disk due to its close projected separation from the inner edge of the outer disk \citep{Mesa2019,Stolker2020,wang2020}. 

Current astrometric measurements of the PDS\,70 planets span a baseline of more than 7 years. Orbital fits to the astrometry place PDS\,70\,b on a slightly eccentric ($\epsilon \sim$0.2) orbit, while the orbit of PDS\,70\,c is consistent with being circular \citep{wang2021}. With semi-major axes of $\sim$21\,au and $\sim$34\,au, respectively, the planets are likely locked in a dynamically stable 2:1 mean-motion resonance \citep{wang2021}, as supported by hydrodynamical simulations of the system \citep{bae2019,toci2020}.

The masses of the two planets, are still uncertain, but have been estimated using a range of approaches, all of which converge on masses that lie solidly in the planetary regime.
Dynamical stability and constraints on disk eccentricity confer upper limits of $\lesssim$ 10 \Mjup on both companions \citep{wang2021,bae2019}. Comparisons of near-IR spectro-photometry to evolutionary and atmospheric models yield masses in the range of $\sim$1 to $\sim$17~\Mjup \ \citep{Keppler2018,Mueller2018,Haffert2019,Mesa2019b,Stolker2020,wang2020}. Measured H$\alpha$ line properties suggest dynamical masses of $\sim$12 and $\sim$11 \Mjup \ for PDS\,70\,b and c, respectively \citep{Hashimoto2020}, consistent with the range of masses derived from near-IR spectro-photometry.

\rev{PDS\,70\, b and c present very red and almost featureless near-IR SEDs compared to other planetary mass companions, hinting at the presence of dust at or near the planets \rev{which may originate in the planetary atmosphere, the accretion column, or in a surrounding circumplanetary disk} \citep{Muller2018,christiaens2019,Stolker2020,wang2021,Cugno2021}.} 
Further evidence of circumplanetary dust around PDS\,70\,c comes from ALMA sub-millimeter continuum observations, revealing compact emission co-located with the H$\alpha$ and near-IR emission of the planet \citep{Isella2019,Benisty2021}. Some emission is detected close to PDS\,70\,b, but it appears faint and with an unclear morphology.  In Figure \ref{fig:pds70}, middle right, and right panels, the sub-millimeter continuum observations are presented with two different angular resolutions. At very high angular resolution (right), the emission co-located with PDS\,70\,c is separated from the outer disk, but the faint emission around PDS\,70\,b is not retrieved. The dust mass of the CPD, as inferred from these continuum observations, is on the order of $\sim$0.01 Earth masses, where the precise value depends on the assumed grain properties \citep{Benisty2021}. It is not clear whether the nature of the CPDs around PDS\,70\,b and PDS\,70\,c is the same (e.g., accretion or decretion disk) considering the fact that one is located in a very gas-depleted  \rev{cavity}, while the other is adjacent the main disk reservoir. 

The presence of dust at or near the planets can also be inferred from optical observations. Following the theoretical models of \cite{Aoyama2018} and neglecting any foreground extinction, the line flux ratio of H$\beta$ to H$\alpha$ originating from the post-shock region of the planetary atmosphere should be near unity. Considering the clear detection of H$\alpha$ but non-detection of H$\beta$, \cite{Hashimoto2020} derived lower limits on the optical extinction of $>$ 2.0 mag and $>$1.1 mag for PDS\,70\, b and c, respectively, assuming that the post-shock region is the primary source of hydrogen line emission. This extinction is presumably caused by the presence of small, unseen dust grains coupled to the gas within the cavity.

Theoretical models predict that the presence of a massive planet in a disk generates a pressure bump outside of its orbit, trapping large dust particles and allowing small grains to flow through the \rev{cavity} \citep{Pinilla2012}. Detailed hydrodynamical simulations by \cite{bae2019} \rev{strongly suggest} that PDS\,70\,b and c open a common \rev{cavity}, trapping submillimeter particles at distances similar to the observed location of the mm continuum ring. 
This is supported by the fact that  disk rotational velocities, as measured from $^{12}$CO observations with ALMA, reveal radially extended perturbations and only recover Keplerian rotation at regions close to the continuum ring \citep{Keppler2019}. This indicates a peak in the radial pressure profile close to the radial accumulation of sub-mm particles, as expected for the process of particle trapping.

\subsection{Protoplanet IR candidates} \label{sec:protoplanets}



Embedded protoplanet candidates have been put forward in a number of disk systems in recent years \citep{Kraus2012a,Quanz2013a,Brittain2013,Biller2014,Reggiani2014,Quanz2015,Sallum2015,Currie2015,Garufi2016,reggiani2018,wagner2019,Gratton2019}, but many have proven difficult to robustly verify as point-sources \citep{Thalmann2016,rameau2017,follette2017,Mendigutia2018,ligi2018,Sissa2018,currie2019}. The complex morphologies of the disks in which protoplanet candidates are embedded \rev{make} disentanglement of planet and disk signals complex, and the fidelity of post-processing algorithms for robust signal detection is a subject of debate. The first reported protoplanet detections was 
of a $\approx$15AU planetary-mass companion to the star LkCa 15, LkCa 15~b, by non-redundant aperture masking (NRM) \citep{2012ApJ...745....5K}. An object consistent with the \citet{2012ApJ...745....5K} companion was detected in several additional NRM epochs, as well as directly in H$\alpha$ emission by \citet{Sallum2015}. The `b' companion, as well as two other point sources, LkCa 15 `c' and `d', were reported in several additional NRM epochs \citep{2016SPIE.9907E..0DS}. However, the coincidence of the LkCa 15 companions with an inner disk component imaged in polarized intensity \citep{Thalmann2016} and total intensity \citep{currie2019} has led to speculation that it may be \rev{due to the scattered light signal from an asymmetrically  disk}. In fact, the candidates LkCa 15 b,c,d, HD100546 b,c, and HD~169142 b,c have been put forward as both protoplanets \citep{2012ApJ...745....5K,Sallum2015,2015ApJ...807...64Q,Currie2015,Biller2014} and potential disk \rev{signatures} \citep{Thalmann2016,currie2019, follette2017, 2017AJ....153..244R, Pohl2017, Biller2014, ligi2018} in the literature. Other protoplanet candidates, such as those in MWC~758 have yet to be fully verified \citep[e.g.][]{reggiani2018}, \rev{with additional observations, to rule out a disk contribution that remains significant enough that it could lead to false detection}. 

\begin{figure*}[!t]
    \centering
    \includegraphics[width=0.93\textwidth]{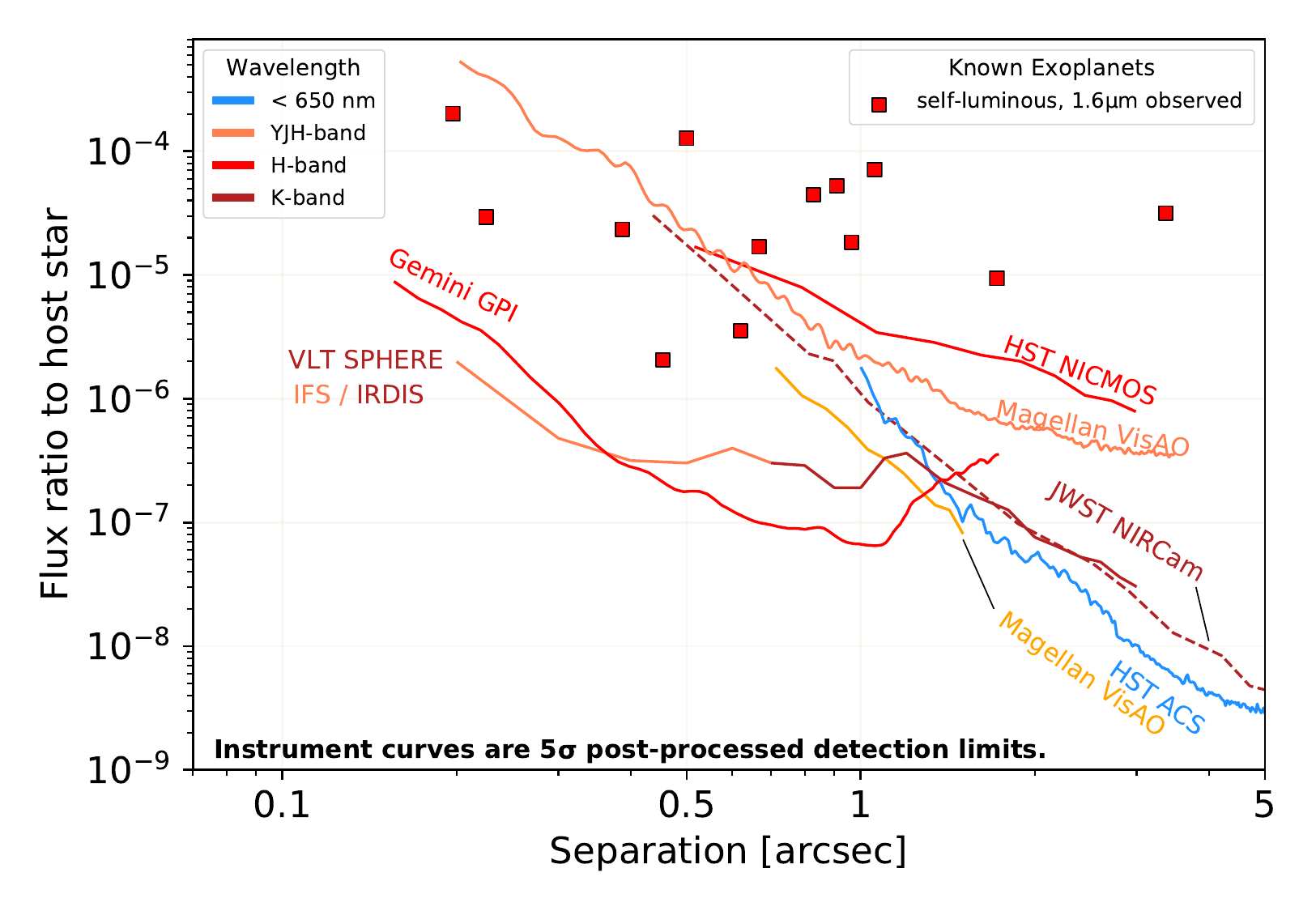}
    \caption{\small Contrast curves at optical and near IR wavelengths from a selection of ground and spaced based high-contrast imaging instruments. The curves depict 5$\sigma$ post-processed contrast limits, and are color coded by wavelength of observation. The points depict the H-band contrasts of directly-imaged companions.  Though the post-processing techniques and exposure times are inhomogeneous, the datasets used to derive them are all exemplary, representing an approximation of the `best' achievable contrasts with each instrument. Image Credit: Vanessa Bailey (\texttt{https://github.com/nasavbailey/DI-flux-ratio-plot}) Sources: STIS handbook; E. Choquet, J. Krist, B. Macintosh, personal communication; \citet{Beichman2010, Vigan2015}}
    
    \label{fig:contrast}
\end{figure*}

\begin{figure}[!t]
    \centering
    \includegraphics[width=0.5\textwidth]{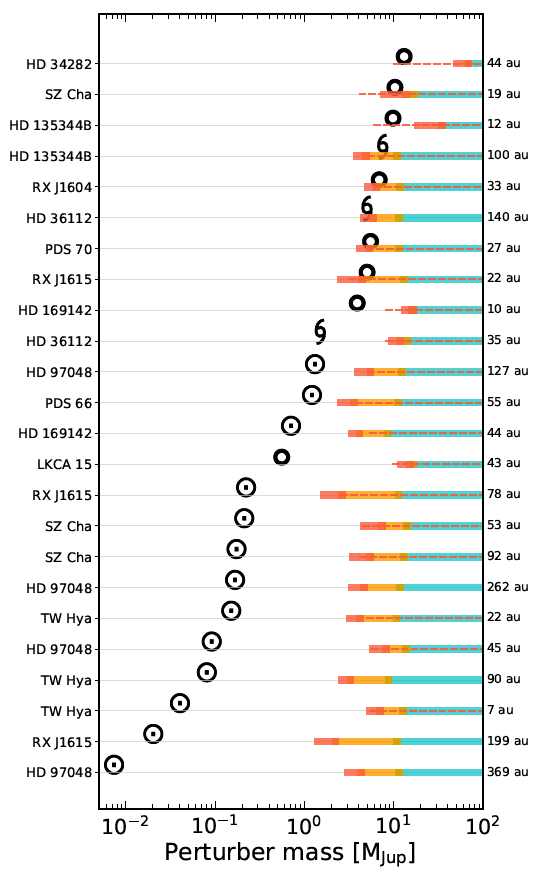}
    \caption{\small 5-$\sigma$ detection limits compared to planets masses expected from modeling of the substructures. The masses are estimated using the AMES-DUSTY, BEXHOT and BEX-WARM initial starts models (red, orange, blue, respectively). \rev{The locations of the substructures for each star are given on the right.}    From \cite{Asensio-Torres2021}.}
    \label{fig:detectlim}
\end{figure}

Although current high-contrast imaging instrument limits \rev{for point sources} are approaching contrasts of nearly 10$^{-5}$ at 0$\farcs$1 and 10$^{-7}$ at 0$\farcs$5 (see Figure \ref{fig:contrast}), these levels are achievable for only a small number of bright, nearby systems. Protoplanets are both farther away \rev{(d$\sim$150-200 pc)} and more heavily extincted (by material local to the system and interstellar dust) than most directly imaged exoplanets \rev{(d$\sim$50 pc)}. Planet searches at longer wavelengths \rev{\citep[e.g.][]{Stone2018,Launhardt2020,Jorquera2021}} are promising avenues for robustly detecting thermal emission from embedded protoplanets. \rev{Depending on the mass and age of the planet, and thus its temperature, the peak of the planet emission is located in the near or mid-IR, maximizing the photons received from the planet and at the same time minimizing the required contrast relative to the central star. Furthermore dust opacities are lower at longer wavelengths (\citealt{Ossenkopf1994, Woitke2016}), allowing for an easier detection of heavily embedded planets. Simulations of the expected fluxes and a discussion of the detectability of embedded planets with surrounding circumplanetary material is given in \cite{Szulagyi2019}. At these longer wavelengths, angular resolution however becomes increasingly challenging. Ground based instrumentation in the L (3.8$\mu m$) and M-bands (4.78$\mu m$), is limited to planets outside of $\sim$125\,mas, which translates into $\sim$20\,au for the nearby star forming regions. Mid-IR ($\sim$10$\mu m$) observations by the James Webb Space Telescope will only have an angular resolution of 425\,mas. This limits searches for embedded planets to the most extended disks at these wavelengths.     }
\rev{Under the assumption that the substructures observed in disks are due to planets, a putative underlying population of protoplanets could  be inferred,} as has been done at longer wavelengths \citep{Zhang2018,Lodato2019}. Using analytical prescriptions for the gap properties, \cite{Asensio-Torres2021} derived the expected planet masses for a number of features in disks observed with SPHERE and compared those to the detection limits under various evolutionary models. The results are shown in Figure \ref{fig:detectlim}, extracted from \cite{Asensio-Torres2021}. We note that this approach, while insightful, remains model-dependent, in particular on the assumptions on the disk properties (viscosity, temperature structure).


\subsection{Search for accretion tracers} \label{sec:protoplanetsHalpha}

Disambiguation of protoplanets from ambient circumstellar disk structures requires clear separation from disk material (as in PDS 70~b), detection of orbital motion, or emission that is inconsistent with circumstellar disk emission, such as accretion signatures. \rev{Excess emission from accreting objects is typically thought to consist of: (a) continuum emission from an accretion ``hot spot" peaking in the NUV--optical range, (b) excess line emission (primarily Hydrogen) from infalling or shocked gas at a range of wavelengths, and (b) excess near-IR emission from the warm accretion disk \citep[e.g.][]{2016ARA&A..54..135H, 2015ApJ...799...16Z}.}

\rev{\paragraph{Continuum excess} in the blue optical to ultraviolet has been successfully detected for several bound brown dwarfs and even one protoplanet, PDS 70 b, with HST \citep{2014ApJ...783L..17Z, Zhou2021}. Low-resolution spectroscopy and spectrophotometry of these sources yields important constraints on continuum emission from the accretion hotspot. These studies are limited to the widest companions because of the small aperture and relatively limited options for PSF subtraction with HST. Near-IR continuum emission has also been detected from several protoplanets \citep[notably PDS70~b and c,][]{Keppler2018,Muller2018}, and protoplanet candidates, including at very tight separation with interferometric techniques \citep[e.g.,][]{Sallum2015,Biller2014}. Some attempts have been made to quantify the contribution of the circumplanetary accretion disk to emission from protoplanets at these wavelengths \citep[e.g.,][]{2015ApJ...799...16Z,2015ApJ...803L...4E}, however uncertainties in accretion disk structure and disk and planetary physical properties make disentangling planetary photospheric and accretion disk contributions difficult.}

\rev{\paragraph{H$\alpha$} is among the brightest accretion emission lines, and computational models suggest that planet/star contrasts at H$\alpha$ should be several orders of magnitude more favorable than thermal emission in the \rev{near-IR} \citep{Mordasini2017}. This allows H$\alpha$ excess to be detected from the ground, where larger telescope apertures combined with shorter wavelength emission allows for detection of accretion from more tightly-separated companions.} 
Achievable contrasts at H$\alpha$ inside of transitional disk cavities are typically $\approx10^{-3}-10^{-2}$, where only the most massive and/or strongly accreting planets are expected to be detectable with current technology \citep{Mordasini2017}, and this detection requires aggressive PSF subtraction. 

Despite the technical complexities, successful detections of accreting protoplanets have allowed for estimation of planetary accretion rates, an important constraint on their evolution, ultimate mass, and perhaps even informing their formation mechanism \citep[e.g.,][]{2015MNRAS.449.3432S}. Initial estimates of accretion rates for protoplanets have been high, comparable to those of T Tauri stars ($\sim10^{-11}--10^{-9}M_{\odot}/yr$). \rev{These estimates should be interpreted as very rough approximations, as calibration of accretion signatures remains difficult. At present, it is unclear to what extent the magnetospheric accretion paradigm for higher mass objects holds at planetary masses.}

 \begin{figure}[t]
 \centering
  \epsscale{1.0}
  \includegraphics[width=0.4\textwidth]{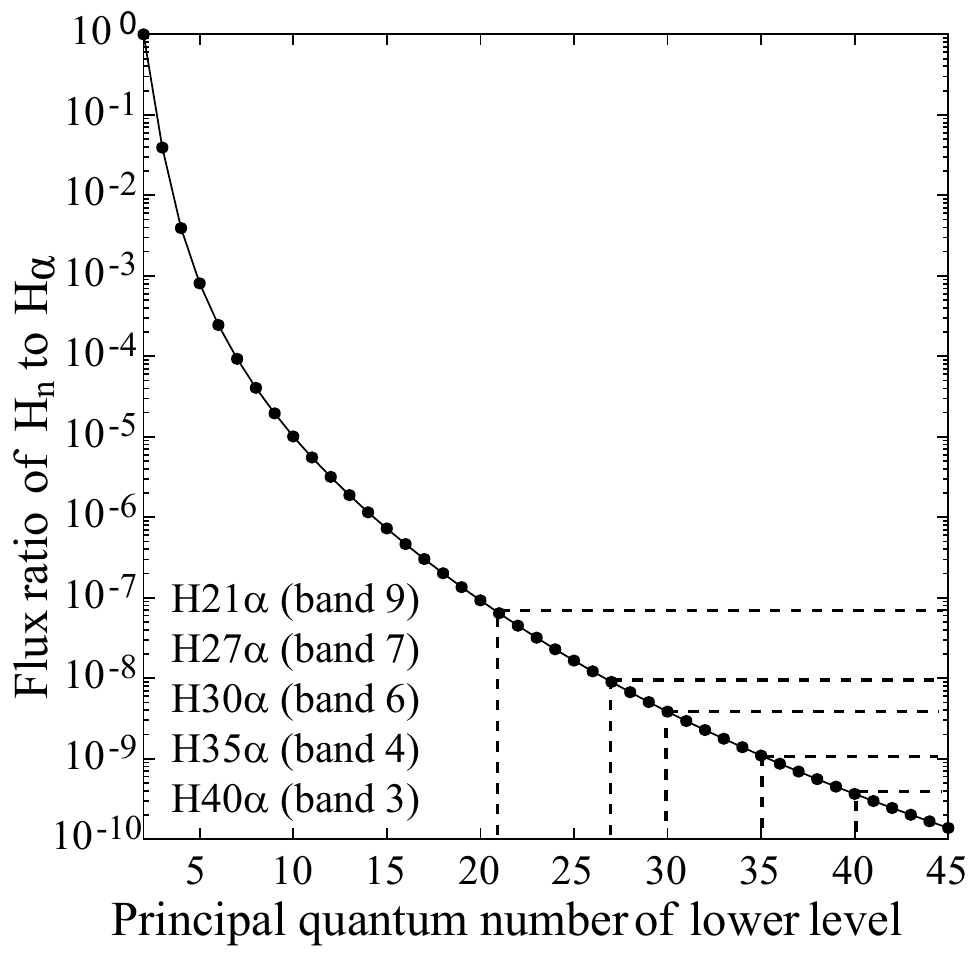}
  \caption{\small Predictions for intensity ratios of Hn$\alpha$ to H$\alpha$ calculated via the one-dimensional radiative hydrodynamic models of \citet{Aoyama2018}. } 
  \label{fig:hrrl_flux}
 \end{figure}

\rev{Conversion of an accreting object's luminosity or contrast at a single accretion-tracing wavelength to a mass accretion rate estimate relies on a number of factors. Key among these are (a) an assumption of a scaling relation between the luminosity at the observed wavelength (e.g., H$\alpha$) and the total accretion luminosity at all wavelengths, and (b) the companion's physical properties, including mass and radius. For classical T-Tauri stars, scaling relations have been established empirically through a range of multiwavelength observations \citep[e.g.,][]{2012A&A...548A..56R,2017A&A...600A..20A}, however it is unclear to what extent these relations hold at lower masses. Poor understanding of magnetic field strengths and geometries for protoplanets make it difficult to infer the precise accretion mechanism (e.g., magnetospheric accretion via a column of infalling gas or more diffuse boundary layer accretion). Furthermore, while accretion line emission for T Tauri stars is believed to originate primarily in the infalling accretion column, the lower temperatures in a protoplanetary atmosphere are predicted to make the post-shock region the dominant source of Hydrogen line emission from protoplanets \citep{Aoyama2018, 2020arXiv201106608A}. The very different physical properties in this region may dramatically alter line-to-total accretion luminosity relations for lower mass objects. Planetary masses and radii are also much more poorly constrained, observationally and theoretically, compared to stars, yielding large uncertainties in mass accretion rate estimates.}


Though properties of gaps, such as width and degree of depletion, have been used to predict planetary masses \citep[e.g.,][]{Bae2018, Zhang2018, Lodato2019}, they do not provide unique constraints due to the degeneracies of planet mass with disk viscosity and aspect ratio \citep{Fung2014a}. Since meridional flow in planetary gaps is predicted to be supersonic, accreting gas shocks in the planetary or circumplanetary disk atmosphere. When the velocity of accreting gas is high (more than $\sim$30~km/s; \citealt{Aoyama2018}), shocks cause the gas temperature to be high enough ($\gtrsim$10$^{4}$~K) to ionize atomic hydrogen, leading to line emission.  Hydrogen radio recombination lines (HRRL) for the higher principal quantum number levels (n $\gtrsim$ 20) might be a good kinematic tracer of accreting protoplanets. At radio wavelengths, dust extinction in the protoplanetary disk is negligible compared to H$\alpha$, especially in the optically thin gap region of the disk. If detected, mm Hydrogen line ratios will allow for estimation of planetary masses, given assumptions about planetary radii and free fall velocities, by applying radiation-hydrodynamic models of the shock-heated accretion flow \citep{Aoyama2018}.

\subsection{Kinematic detections of protoplanets} \label{sec:kinks}


The velocity structure of disk gas as revealed by ALMA provides a unique, indirect signature of protoplanets (see Chapter by \textit{Pinte et~al.}). A massive planet embedded in the gas disk is expected to cause a local deviation from Keplerian velocity field \citep[e.g.,][]{Perez2015}, \rev{with a characteristic pattern of sub- and super-Keplerian velocities near the planetary perturber as a result of the lower gas pressure at the location of the planetary clearing  \citep[e.g.,][]{Kley2001,Tanigawa2012}.}
\rev{This can be traced in the gas velocity channels as seen in Figure\,\ref{fig:alma} (right; so-called velocity kinks) and also as a doppler-flip signature \citep{Casassus+Perez2019} in velocity maps}. 
\rev{Recently, \citet{Izquierdo2021} proposed a statistical method to identify and quantify kinematical perturbations and considered the contributions from both the upper and lower emitting disk surfaces. This framework was validated with observational data on HD163296 \citep{Izquierdo2021b}.}

These deviations depend on the planet masses, and can be used to estimate the unseen planet mass by comparing with hydrodynamic simulations. To date roughly ten objects have been reported to possess the so-called velocity kink structures, although some are tentative detections at low signal-to-noise ratio. Table~\ref{tab:kink} summarizes their properties, corresponding planet mass estimates, and limits on planet masses at those locations from near-IR imagery. 

Another \rev{promising} kinematic detection method for protoplanets \rev{would be} the presence of a `meridional flow'. The gas flows from the disk onto the planet through the pole and produces a 3D velocity signature that could be clearly mapped in HD163296 \citep{Teague2019a} and HD169142 \citep{Yu2021}. Hydrodynamic simulations also predict a kinematic signature for circumplanetary disks in the form of a point source in the velocity channel maps \citep{Perez2015}. A tentative detection of such a point source was reported for PDS~70, \rev{using ALMA observations of the $^{12}$CO (3–2) line}, in \citet{Keppler2019}. 

    \begin{table}[!t]
    \begin{center}
      \caption{Planet mass estimation from ALMA kinks compared to near-IR detection limit}\label{tab:kink}
      \begin{tabular}{lcccc}
      \\
      \hline \noalign {\smallskip}
        Object & Radial   & Expected & Detection  & Refs \\
               & location & mass     & limit      &      \\
               & (au)     & ($M_{\rm Jup}$) & ($M_{\rm Jup}$) &  \\
       \hline \noalign {\smallskip}
        HD 163296 & 260 & 2    & 2      & 1,2 \\
                  & 68  & 1$-$3 & 5      & 3,2 \\
        DoAr 25   & 60  & 1$-$3 & 5      & 3,4 \\
        IM Lup    & 110 & 1$-$3 & 2      & 3,5 \\
     Elias 2$-$27 & 37  & 1$-$3 & 20     & 3,5 \\
        GW Lup    & 61  & 1$-$3 & 4      & 3,6 \\
        HD 143006 & 23  & 1$-$3 & $>$20  & 3,6 \\
        Sz 129    & 47  & 1$-$3 & $>$20  & 3,6 \\
        WaOph 6   & 63  & 1$-$3 & 3      & 3,6 \\
        HD 97048  & 130 & 2$-$3 & 2      & 7,8 \\
        HD 100546 & 26  & 5$-$10 & $>$20 & 9 \\
       \hline \noalign {\smallskip}
      \end{tabular}
      \begin{tablenotes}
References for respectively kink's radial locations, expected planet masses, and detection limit at the kink's locations: (1) \citet{Pinte2018}, (2) \citet{Guidi2018}, (3) \citet{Pinte2020}, (4) \citet{Uyama2017a}, (5) \citet{Launhardt2020}, (6) \citet{Jorquera2021}, (7) \citet{Pinte2019}, (8) \citet{Ginski2016}, (9) \citet{Perez2020}. 
\end{tablenotes}
\end{center}
  \end{table}

\section{Future prospects} \label{sec:future}

\subsection{Perspectives}
As mentioned before, multi-wavelength observations trace different vertical layers in the disk: optical/near-IR wavelengths \rev{trace} small grains in the upper surface layer, while (sub-)millimeter wavelengths \rev{trace}  larger pebbles \rev{located at deeper layers within the disk, closer to} the disk midplane. One of remaining questions of our field concerns the connection between the observed structures in these wavelengths. Some objects exhibit very different sub-structures in different wavelengths as showed in Figure\,\ref{fig:alma} (e.g., AB~Aur in \citealp{Fukagawa2004,Tang2017}; SAO~206462 in \citealp{Muto2012,Cazzoletti2018b}). 
The ALMA gas observations trace the optical depth $\tau\sim$1 layer that corresponds to the gas isotope abundances (e.g., $^{12}$CO/$^{13}$CO=67, $^{12}$CO/C$^{18}$O=444, C$^{18}$O/C$^{17}$O=3.8, \citealp{Qi2011}), with less abundant gases such as C$^{17}$O tracing deeper disk layers. Multi-wavelength observations combined with radiative transfer modeling, could serve as building 3D structures of the disk and help understanding of possible common origin of different substructures in different wavelengths. 

Near-IR observations of the polarization fraction in linear polarization could reveal the dust porosity in the disk. Since planets are thought to form via collisional aggregation of dust grains \citep[e.g.,][]{Weidenschilling1993,Dominik2007}, overcoming bouncing barrier \citep[e.g.,][]{Wada2011} and radial drift barrier \citep[e.g.,][]{Okuzumi2012} is an important step in grain growth. Dust porosity could be crucial to overcome these barriers. Numerical simulations by \citet{tazaki2019} suggest that high polarization fraction ($\sim$65--75\%) in linear polarization is attributed to highly porous dust aggregates, while more compact dust structure tends to show low polarization fraction ($\sim$30\%). In observed near-IR polarization fractions, both of high ($\sim$50\%) polarization fraction \citep[e.g.,][]{Silber2000,Perrin2009,Tanii2012,Poteet2018} and low ($\sim$30\%) one \citep[e.g.,][]{Canovas2013,Avenhaus2014a} have been reported in literature. Thus, some of these observations may suggest the presence of large porous aggregates. Future ExAO observations of polarization fraction would provide more reliable results.

In addition, optical/near-IR circular polarization observations of protoplanetary disks might provide useful information of magnetic fields. Magnetic fields are thought to play an important role in the disk evolution via magnetically driven disk winds \citep[e.g.,][]{Suzuki2009,Suzuki2016}. The spatial distribution of close-in super-Earths is attributed to the suppression of rapid inward migration (i.e., type~I migration) by magnetically driven disk winds \citep{Ogihara2018a}, and thus, observations of the magnetic field strength and structure would be valuable to understand formation of close-in super-Earths. Circular polarization has been observed in circumstellar materials around young stars \citep[e.g.,][]{Fukue2010,Kwon2013,Kwon2014,Kwon2016a,Kwon2016b,Kwon2018}, and has been interpreted as a result of scattering from aligned non-spherical dust grains \citep[e.g.,][]{Gledhill2000,Wolf2002} and the dichroic extinction \citep[e.g.,][]{Lucas2005}. Non-spherical dust grains could be aligned with the local magnetic field \citep[e.g.,][]{Davis1951,Dolginov1976,Draine1996,Draine1997,Andersson2015}. \citet{Tazaki2017} expect that circular polarization in protoplanetary disks can be produced by scattering from aligned dust grains. Future observations of circular polarization with VLT/SPHERE and Subaru/SCExAO would constrain not only dust properties (i.e., spherical or non-spherical) but also the magnetic field in disks.


The advances in instrumentation in the last decade have led to a high temporal stability of ground based facilities, which was prior only achievable with space based observations. This development has opened up the time domain to be included in studies employing resolved scattered light imaging. This allows to trace the proper motions of sub-structures, such as spiral arms or dust-clumps, but also links the resolved observations to the inner, un-resolved disk regions via the variability of shadow features.

 \begin{figure*}[t]
  \includegraphics[width=1.0\textwidth]{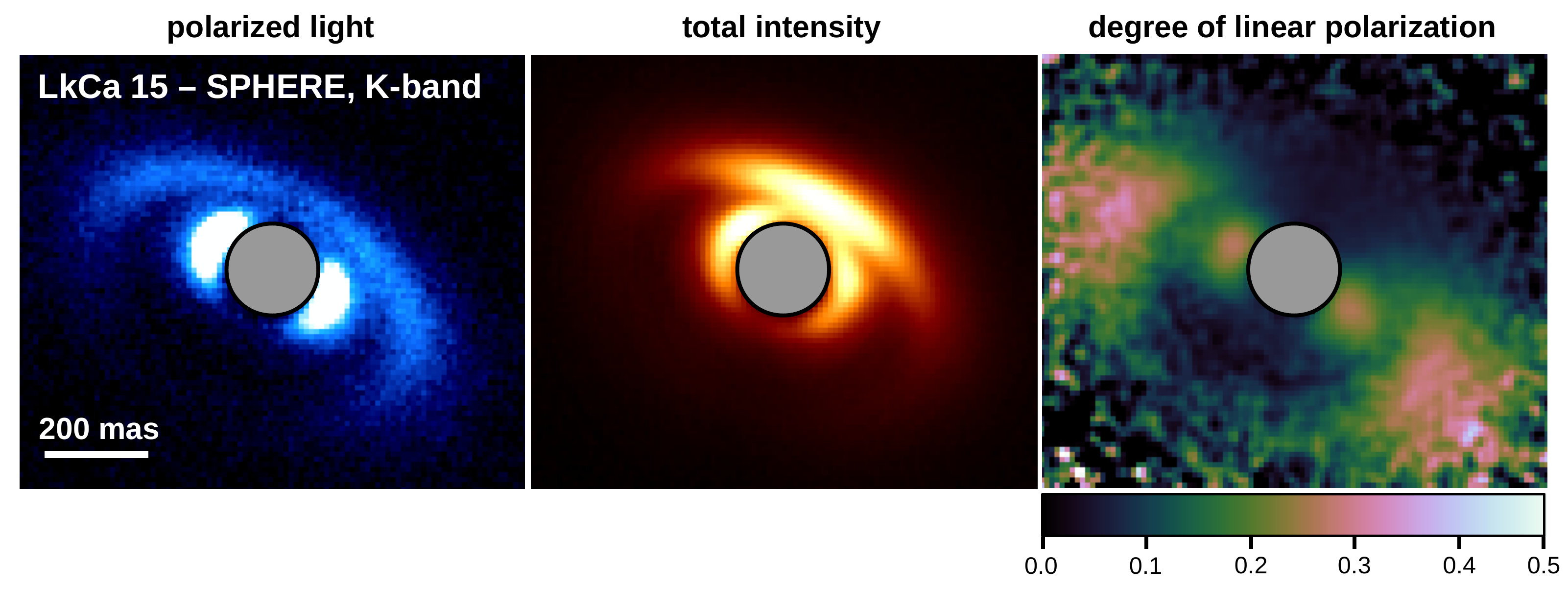}
  \caption{\small Recent VLT/SPHERE observation of the LkCa\,15 system using a combination of the polarimetric imaging and the "star-hopping" mode for advanced reference differential imaging. The left panel shows polarized intensity and the middle panel shows total intensity. Both are displayed on a linear stretch of the color map which is a factor 10 larger for total intensity. The right panel shows the derived degree of linear polarization computed from the polarized intensity and total intensity image.}
  \label{fig:lkca15-dolp}
 \end{figure*}

The proper motion of substructures gives insight into the mechanisms shaping the disk. 
\rev{For example,} the rotation of spiral features has been the focus of attention. The angular velocity of the spirals can in this case be connected to the orbital distance of the perturbing body, that launches the spiral density wave in the gas. \cite{Ren2018} used a combination of HST/NICMOS, Keck/NIRC2 and VLT/SPHERE data to investigate the relative position of the spiral arms in MWC\,758 over a time frame of more than 10\,yr. They found that their measurements were best fit by a pattern speed of $0.6^{+3.3}_{-0.6}{}\,\mbox{deg}\, {\mbox{yr}}^{-1}$. With five additional years of monitoring by VLT/SPHERE, they were eventually able to constrain this to 0$^\circ$.22$\pm$0$^\circ$.03 (\citealt{Ren2020}), corresponding to a planetary perturber at an orbit of $\sim$172\,au.  A similar pattern speed was found by \cite{Boccaletti2021}, from their VLT/SPHERE data. 
This series of studies demonstrates the advancements in temporal studies enabled by modern ground based instrumentation, with a dedicated astrometric calibration strategy (see e.g. \citealt{Maire2016} for VLT/SPHERE).
With growing time baseline we will in the future be able to trace the motion of most spiral features. The pattern speed, combined with the direction of motion (leading versus trailing spirals), will enable us to constrain the perturbing mechanism. Given the astrometric accuracy of currently available instrumentation, a motion such as in the spiral arms around MWC\,758 can be detected within roughly two years. 

As we discuss in section~\ref{sec:shadows}, \rev{azimuthal} shadows are a common feature seen in scattered light observations. They enable us to draw conclusion on the inner disk geometry (typically inside of 10\,au), which is inaccessible to current imaging observations, due to limited angular resolution. However, several studies have now found that the shadowing signatures on the outer disks are changing over time (\citealt{Debes2017, Stolker2016a, Stolker2017, Pinilla2018b}). The timescales of some of these variations are as short as days, suggesting that dynamical processes in the innermost disk regions are responsible. Suggested processes include the precession of an inner disk due to a perturbing short-period planet or episodic accretion. These first studies motivate further monitoring in the future, in particular to detect periodicity and the shortest timescales of variation. 
Such monitoring campaigns may in the future be combined with near-IR interferometry, using the latest generation of instruments such as VLT/GRAVITY or VLT/MATISSE, to access the innermost disk regions.

\subsection{Instrument upgrades and new facilities} \label{sec:upgrade}

Multiple avenues for further improving optical and IR imaging of circumstellar disks and embedded planets are on the horizon. The main dimensions for progress in observed systems are in total intensity calibration, wavelength coverage, contrast limits,   and ultimately higher angular resolution (and smaller inner working angle) with the next-generation of Extremely Large Telescopes (ELTs). In addition to these gains, an important step forward will be to shift the optical brightness and thus stellar mass limitations of our current sample to the bulk of the low-mass star population which is already targeted by ALMA sub-mm surveys.

The vast majority of the high-contrast and high-fidelity disk imaging has been done in linear polarized intensity (PI, polarized intensity = $Q^2+U^2$) and often in only one or two passbands.  The full potential of probing the dust size distribution in disks, a key diagnostic of planet formation processes, can only be unlocked with broad wavelength coverage and with PI images {\em calibrated with total intensity images}. While many ''tricks'' are effective in the searching for point sources (i.e., exoplanets) in messy total intensity imaging -- through angular differential imaging (ADI), using chromatic properties of speckles, and/or use of reference image libraries (e.g., LOCI) -- these methods fail for extended low surface brightness objects such as disks.  The only path to improvement is through improved adaptive optics systems and/or {\em more stable and reproducible} performance stability to allow the stellar halo to be subtracted via either forward modeling or using contemporaneous observations of a reference source. While this method has been attempted for decades, there was been recent progress demonstrated using VLT/SPHERE.  Figure\,\ref{fig:lkca15-dolp} presents polarized intensity and total intensity images of LkCa15, obtained simultaneously using the SPHERE star-hopping mode \rev{(\citealt{Wahhaj2021})} that allows \rev{interspersed observations of science target and a reference star with similar spectral properties. This rapid switching combined with the temporal stability of the adaptive optics system enables to obtain a near perfect library of reference star images, matching the scientific target. Combined with  iterative post processing techniques that limit signal over-subtraction, 
this allows for significant improvements to the reference differential imaging technique.  Obtaining both, the polarized and total intensity images, makes it possible to derive the degree of linear polarization, and further characterize the dust grains properties.} 

Other AO systems are also undergoing wavefront sensor upgrades, e.g., GPI2 \citep{Chilcote2020}, SPHERE$+$ \citep{Boccaletti2020b}, SCExAO \citep{Lozi2020}, and MagAO-X \citep{Males2020} in order to improve wavefront stability, peak strehl, and natural guide star sensitivity (especially, in the near-IR). In particular the last point, will unlock the bulk of the low-mass ($<$0.5\,M$_\odot$) stellar population in nearby star forming regions for observations, i.e. the most common members with the most typical compact disks. Embedded protoplanet studies will benefit from the same instrument upgrades, as well as the advancement of mid-to-high-resolution spectroscopic high-contrast imaging instruments such as MUSE and GRAVITY.

Even with PI and total-I imaging, the secrets of the dust populations require {\em multi-wavelength} observations.  Optical to near-IR wavelengths probe the characteristic dust sizes in the upper layers and can reveal important grain characteristics, including non-trivial properties arising from agglomerates rather than perfect Mie-scattering spheres.  The scattering phase function can also be derived for disks with well-understood geometries, another probe of dust grain structure.  Most protoplanetary disks lack the detailed phase-function work we have seen for some debris disks \citep[e.g., HR4796][]{Arriaga2020}. We might expect radial-, disk-height-, and age-dependence during the planet formation process and this represents a topic with immense potential.   In addition to improved optical/near-IR coverage with existing AO instruments, mid-IR disk imaging from VLT/ERIS, VLT/CRIRES$+$, VLTI/MATISSE will play an increasing role in disk studies.

Lastly, we look beyond the current instruments to the next-generation of telescopes.  The James Webb Space Telescope will offer new high-contrast views of disks in the IR, although with a limited inner working angle.  A stable point spread function should allow subtraction of the central star to reveal new details in nearby disks. Its sensitivity in near-IR atmospheric windows and in the MIR will enable studies of new and more sensitive accretion diagnostics. The Roman Space Telescope  will have an experimental coronagraph operating in the optical that, while designed primarily for exoplanet science, will nonetheless offer an unprecedented view of circumstellar disks in the optical \citep{Poberezhskiy2021} and will acheive H$\alpha$ contrasts well below current ground-based limits \citep{Mennesson2018}.  We hope for detection of the exoplanets that have been predicted for some disks based on analysis of the dust substructures.   Beyond the relatively-small aperture space telescopes, we can look forward to the 30-m class telescopes under construction on Earth.  While no polarization capabilities are under development for the first generation of ELT instruments, we expect the $\sim2-4\times$ boost in spatial resolution will allow important disk imaging with the following near-IR AO imagers:  GMT/IRS, TMT/IRIS,  and ELT/MICADO.   Another exciting ELT instrument for disks is the mid-IR instrument ELT/METIS -- bringing 50-milliarcsecond angular resolution to the $10\mu$m N band. The increased sensitivity and angular resolution  of these instruments will allow direct detections of many young exoplanets (as well as proper motion studies) in nearby disks, finally revealing the links between planet formation and disk substructures such as spirals, rings, etc. Polarimetry has been mentioned for inclusion in some 2nd-generation, extreme AO instruments (e.g., TMT/PSI, ELT/PCS) but will likely not be available until $>$2035.

\section{Summary} 
 \label{sec:summary}

The years since PPVI have seen a revolution in the field of high spatial resolution, high contrast imaging of planet-forming disks and protoplanets. In addition to what ALMA has been able to achieve, direct imaging at optical and near-IR wavelengths have opened up a separate, complementary view.  At the time of PPVI, we had the HST images of only a few planet-forming disks, and the Subaru telescope with the SEEDS program along with VLT/NACO had started to open up this space.
While HST is still ruling supreme in terms of sensitivity at large working angles, the advancement of adaptive optics, advanced coronagraphy, and modern contrast-enhancing techniques on 8-10 m ground-based telescopes have made it possible to achieve detailed imaging at much smaller working angles. In the present chapter, we have described the outcome of this revolution. We summarize the key points of our review below.

\begin{enumerate}

\item The key contrast-enhancing technique has been Polarization Differential Imaging (PDI), allowing a very clean separation of disk photons from stellar (halo) photons without self-subtraction and the clearest detections of disk structures.  Reference Differential Imaging allows access to total intensity, while Angular Differential Imaging drills down onto point sources, and Spectral Differential Imaging starts to give access to accretion-related line emission.


\item PDI observations allowed us to routinely image the disk portion from 15 au to 100 au that is hardly accessible by HST. Space observations do sometimes show a larger extent than that of the detectable polarized light, as also corroborated by the large size of gas disks seen with ALMA.

\item The overall flaring disk shape predicted in theoretical models for disks in hydrostatic equilibrium is commonly seen, and the shape of the disk surface can be measured if the disk has either ring-shaped features or a visible sharp outer edge.

\item The brightness of disks can vary significantly as a function of illumination geometry and grain properties, covering a range of at least a factor of 100 in luminosity-corrected brightness.  The brightest disks have a dust-depleted cavity, leading to an enhanced illumination of the surface and inner rim of the remaining disk.

\item Measured scattering phase functions point to the existence of dust grains or more likely dust aggregates with sizes of several micrometers in the disk scattering surface.  Some dust is still covered by water and possibly other ices.

\item Practically all bright and extended disks show a variety of structures: cavities, rings, spiral arms, both narrow and broad shadow features.  Multiple rings are detected in many disks, with planets as origin speculated but unfortunately not yet confirmed. Spirals are more frequently found in early-type star disks, can be very prominent and sometimes be traced back to the interaction with a companion. Flocculent spirals which may result from gravitational instabilities or a multiple system in the disk are common as well. Narrow shadow lanes are traced back to highly inclined inner disks, while broad shadow features are probably related to disk warps of only a few degrees. Such misalignments and warps can be caused by inclined companions in the system, but also by late infall onto the disk, as evidenced in the observation of a number of disks.  Some shadow features are variable in time, pointing to the role of surface height variations in the shadowing disk components.

\item Planetary or sub-stellar companions in disks can be found with imaging techniques, looking for the continuum emission of the object, line or continuum emission of a circumplanetary disk, kinematic disturbances in disk gas motions, or accretion-related emission lines.  Just the detection of the PDS70 system with 2 directly detected planets has revolutionized the field.  Many candidates still await confirmation and characterization, and an increase of the number of directly detected planets is arguably the most important goal for further research.
\end{enumerate}

In the future, exceptional power will come from combining multiwavelength high angular resolution observations including scattered light, dust continuum and molecular line observations in similar detail on the same disks, in order to derive the strongest constraints on the properties and physical processes in disks, helping to decipher the environment in which planets for and planetary systems are shaped.

\section*{Willy Kley\protect\newline A tribute to a much-valued friend and colleague} 
 \label{sec:willykley}
 \vspace*{4mm}
\begin{center}
\includegraphics[width=0.485\textwidth]{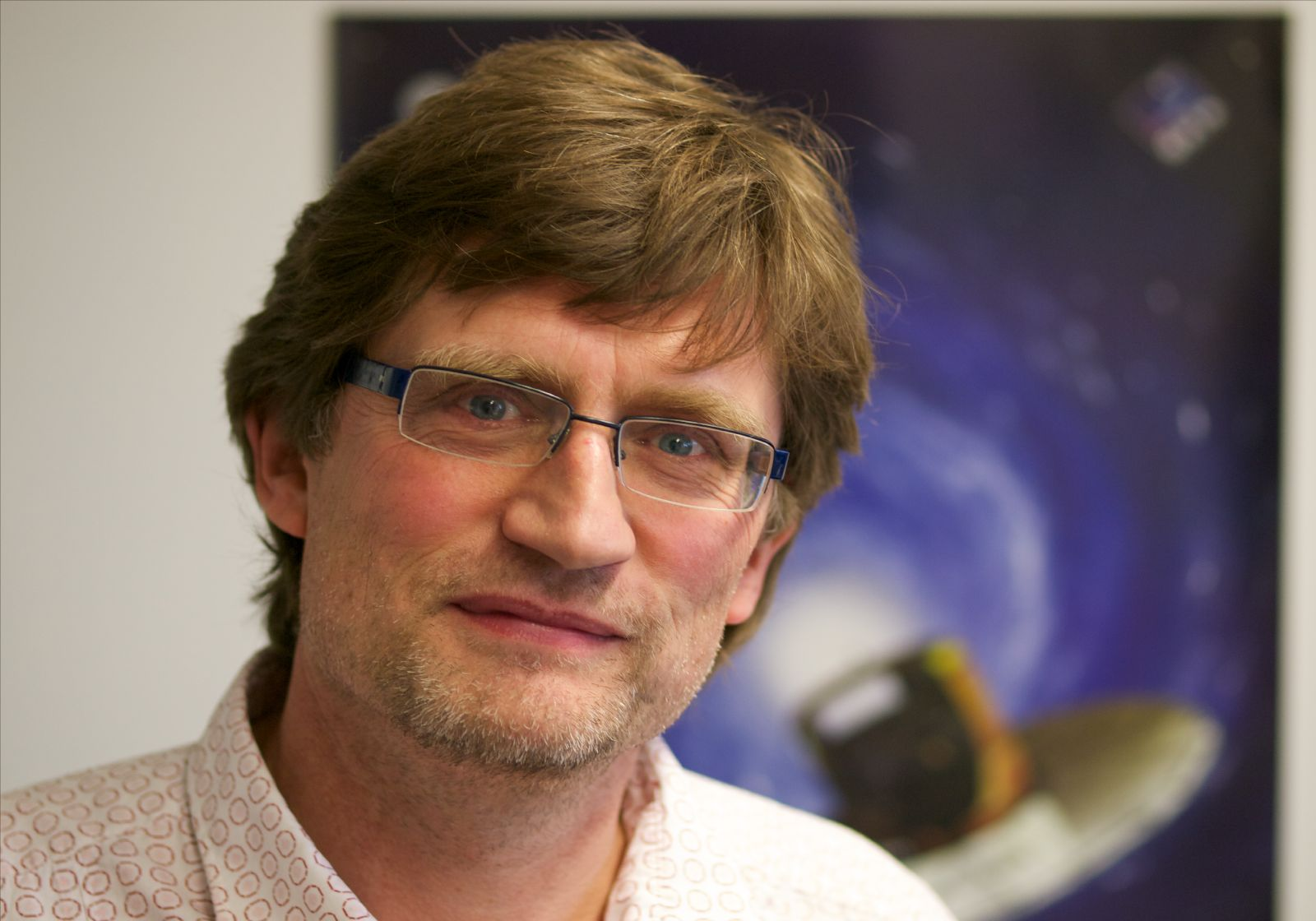}
\end{center}

\vspace*{4mm}
We have all lived through some difficult times during the last two years, but for his friends and colleagues one of the darkest days arrived on 21$^{\rm st}$ December 2021 with the sad and unexpected news of Willy Kley's passing.

Willy will be missed by the science community for many reasons. For his contributions to our understanding of planet formation, and his pioneering work on disc-planet interactions. For his mentorship of students and early career researchers, many of whom are now leading researchers in the field, and for his exemplary commitment to university teaching. For his collegiality, generosity and fundamental decency as a human being.

Willy began his career in the late 1980s as a PhD student in Munich, where, with Gerhardt Hensler,  he introduced the use of hydrodynamical simulations to study accretion disc boundary layers in Cataclysmic Variables. Even at this early stage his papers exemplified the qualities that would become a hallmark of his later publications in planet formation: strongly motivated studies that address key questions in the field; development and deployment state-of-the-art numerical methods; results that lead to new insights, presented with a clear exposition.  

Willy started to work on protoplanetary discs (PPDs) in the early 1990s when he was a postdoc in Santa Cruz working with Doug Lin. He and Doug visited London for a year in 1992, and collaborated with John Papaloizou on various problems in accretion disc theory, especially the role of convection in transporting angular momentum. Over the coming years Willy also published results on the hydrodynamics of boundary layers in PPDs and on FU Orionis outbursts.

The discovery of the first hot Jupiter, 51 Peg b, in 1995 changed everything, and the most important question of the moment became how do such systems actually form? Willy's expertise in PPDs and his finely-honed numerical skills, combined with his instinct for knowing which problems to work on, placed him in pole position to make an immediate impact, which he did with the publication in 1999 of one of his most influential papers \emph{Mass flow and accretion through gaps in accretion discs} \citep{1999MNRAS.303..696K}. For the next 20 years Willy would remain at the forefront of developments that have enhanced our understanding of planet formation and the role of disc-planet interactions.

Willy's scientific contributions, undertaken with many collaborators, are too numerous to detail, but highlights include: the first hydrodynamical study of a multi-planet system embedded in a PPD, followed up by later works on resonance capture to explain the GJ876 multi-planet system; pioneering studies of planet formation and evolution in close binary systems; ground-breaking studies of orbital evolution and gas accretion onto planets embedded in PPDs, employing state-of-the-art numerical methods, realistic equations of state and radiative processes; 3D radiation-hydrodynamic simulations of the Vertical Shear Instability and the settling/mixing of dust in the outer regions of PPDs.
In addition, he contributed significantly to Protostar \& Planets review articles, including those published as part of the PPV, PPVI and now PPVII conferences. In testament to his standing in the field, he was invited in 2012 to lead the publication of an Annual Reviews article on Disc-Planet Interaction and Orbital Evolution, which remains as a standard reference text on the topic. 

My personal recollections of Willy go back to 1992 when he and Doug Lin visited Queen Mary University of London (QMUL) where I was as a PhD student. I had the impossible task of running star formation simulations on the ageing computers we had in the department at QMUL. Thankfully, Willy and Doug had brought with them the latest IBM workstation from Santa Cruz, along with a substantial allocation of time at the San Diego supercomputing centre. As soon as Willy realised my situation, and without prompting or hesitation, he gave me access to both of these resources, literally enabling me to complete my PhD in the process. This act of generosity was the start of a life-long friendship. Willy and I started collaborating scientifically in the late 1990s, by which time we were both working on the problem of disc-planet interactions. Once again I was in receipt of Willy's generosity when he offered to share his simulation code, which was far superior to the method I was using at the time. Working with Willy was always a pleasure, and I recall with great fondness the many wide-ranging discussions we had in restaurants and bars at conferences, or over weissbier and schweinshaxe during my numerous visits to T\"ubingen. He had many qualities as a collaborator and a scientist, but the one that stands out most of all was his love of science and dedication to understanding the world we live in.

Outside of work Willy was a loving and committed husband and father, and is survived by his wife, Martina, and his daughters Lily, Xenia and Sian. He will be missed but not forgotten.

\vspace{5mm}

\noindent Richard P. Nelson\\
\noindent March 10, 2022 

\vskip 5cm 

\noindent\textbf{Acknowledgments} 
We are very grateful to Evan Rich for his help with the GPI datasets and the uniformization of the FITS format for direct imaging data, and to Ryo Tazaki for his help in shaping Section \ref{sec:grains} and for providing Figure~\ref{fig:phase_functions}. We thank Jaehan Bae for fruitful discussions about substructures while writing the review.  We acknowledge undergraduate researchers Alexander DelFranco and Dane Mansfield of Amherst College for their contributions to the design of Figure \ref{fig:gallery}, and Vanessa Bailey for her public code used to generate Figure \ref{fig:contrast}. We thank the three reviewers whose extensive reviews, detailed comments and constructive suggestions helped improve the manuscript. We acknowledge the essential contribution of scientists, engineers, astronomers, telescope operators, working at the VLT, Subaru and Gemini facilities, and for the HST, who allow the community to obtain the results presented in this review.  This project has received funding from the European Research Council (ERC) under the European Union’s Horizon 2020 research and innovation programme (PROTOPLANETS, grant agreement No. 101002188). JDM acknowledges support from NSF/AST-1830728, and KBF from NSF/AST-2009816. C.G. and C.D acknowledge funding from the Netherlands Organisation for Scientific Research (NWO), project number 614.001.552.

\bigskip

\bibliographystyle{pp7}
\bibliography{pp7}

\begin{appendix}
\section{Literature references for gallery figures}
The data sets shown in Figures \ref{fig:gallery}, \ref{fig:brightness} and \ref{fig:spiral-gallery},  were gathered from the archives and reduced with latest pipeline versions of the respective instruments. Here we list the publications in which the data sets were first described. 

\paragraph{References for Figure\,\ref{fig:gallery}}
\label{App: ref: main gallery}

\noindent From upper left to lower right:\\
{\bf Disks with rings:}
TW\,Hya (\citealt{vanBoekel2017}), RXJ1615-3255 (\citealt{Avenhaus2018}), HD\,169142 (\citealt{Momose2015}), Oph IRS\,48 (\citealt{Follette2015}).\\
{\bf Disks with spirals:} GQ\,Lup (\citealt{vanHolstein2021}), EM*\,SR\,21 (\citealt{MuroArena2020}), HD\,34700 (\citealt{Monnier2017}), HD\,135344B (\citealt{Stolker2016a}).\\
{\bf Disks with broad shadows:} EX\,Lup (\citealt{Rigliaco2020}), WRAY\,15-788 (\citealt{Bohn2019}, Wolff et al., in prep.), HD\,143006 (\citealt{Benisty2018}), HD\,139614 (\citealt{Laws2020}).\\
{\bf Disks with narrow shadows:} GG\,Tau (\citealt{keppler2020}), RXJ\,1604.3 (\citealt{Pinilla2018b}), HD\,142527 (\citealt{Hunziker2021}), HD\,100453 (\citealt{Benisty2017}).\\
{\bf Disks with visible backside:} IM\,Lup (\citealt{Avenhaus2018}), MY\,Lup (\citealt{Avenhaus2018}), PDS\,453 (Menard et al., in prep.), HD\,34282 (\citealt{deBoer2021}).\\
{\bf Disks interacting with ambient material:} DO\,Tau (Huang et al., subm.), 2MASS\,J1615-1921 (\citealt{Garufi2020a}), GW\,Ori (\citealt{Kraus2020}), MWC\,758 (\citealt{Laws2020}).

\paragraph{References for Figure\,\ref{fig:brightness}}
\label{App: ref: faint gallery}

From left to right: HK\,Lup (\citealt{Garufi2020a}), J1609-1908 (\citealt{Garufi2020a}), SZ\,111 (unpublished), DoAr\,44 (\citealt{Avenhaus2018}), RY\,Lup (\citealt{Langlois2018})

\paragraph{References for Figure \ref{fig:spiral-gallery}}
\label{App: ref: spiral gallery}

From upper left to lower right:\\
{\bf Symmetric/ double-arm disks:} DZ\,Cha (\citealt{Canovas2018}), V1247\,Ori (\citealt{Ohta2016}), MWC\,758 (\citealt{Benisty2015}), LkHa\,330 (\citealt{uyama2018}), EM*\,SR\,21 (\citealt{MuroArena2020}), HD\,135344B (\citealt{Stolker2016a}).\\
{\bf Asymmetric/ multi-arm disks:} CQ\,Tau (\citealt{Uyama2020}), HD\,100546 (\citealt{Sissa2018}), AT\,Pyx (\citealt{Ginski2022}), HD\,142527 (\citealt{Hunziker2021}), RY\,Lup (\citealt{Langlois2018}), HD\,34282 (\citealt{deBoer2021}).\\
{\bf Flocculent spirals:} SU\,Aur (\citealt{Ginski2021}), GG\,Tau (\citealt{keppler2020}), HD\,34700 (\citealt{Monnier2017}).\\
{\bf Interaction with environment:} WW\,Cha (\citealt{Garufi2020a}), DR\,Tau (\citealt{Mesa2022}), AB\,Aur (\citealt{Boccaletti2021}).\\
{\bf Disks with outer companions:} HD\,100453 (\citealt{Benisty2017}), GQ\,Lup (\citealt{vanHolstein2021}), UX\,Tau (\citealt{Menard2020})

\end{appendix}

\end{document}